\documentclass[prb,twocolumn,showpacs,superscriptaddress]{revtex4-1}
\usepackage{epsfig}
\usepackage{amsmath,amssymb}

\begin{document}

\title{A renormalization-group analysis of the interacting resonant level model
  at finite bias: Generic analytic study of static properties and quench
  dynamics} \author{S. Andergassen,$^1$ M. Pletyukhov,$^1$ D. Schuricht,$^1$
  H. Schoeller,$^1$ and L. Borda$^{2,3}$  \\
  {\small\em $^1$Institut f\"ur Theorie der Statistischen Physik, RWTH Aachen,
    52056 Aachen, Germany} \\
  {\small\em and JARA-Fundamentals of Future Information Technology} \\
  {\small\em $^2$Physikalisches Institut, Universit\"at Bonn, D-53115 Bonn,
    Germany}\\
  {\small\em $^3$Department of Theoretical Physics and Condensed Matter
    Research Group of the HAS\\
    TU Budapest, H-1111 Budapest, Hungary} } \date{\small\today}
\begin{abstract}
  Using a real-time renormalization group method we study the minimal model of
  a quantum dot dominated by charge fluctuations, the two-lead interacting
  resonant level model, at finite bias voltage. We develop a set of RG
  equations to treat the case of weak and strong charge fluctuations, together
  with the determination of power-law exponents up to second order in the
  Coulomb interaction.  We derive analytic expressions for the charge
  susceptibility, the steady-state current and the conductance in the
  situation of arbitrary system parameters, in particular away from the
  particle-hole symmetric point and for asymmetric Coulomb interactions.  In
  the generic asymmetric situation we find that power laws can be observed for
  the current only as function of the level position (gate voltage) but not as
  function of the voltage. Furthermore, we study the quench dynamics after
  sudden switch-on of the level-lead couplings. The time evolution of the dot
  occupation and current is governed by exponential relaxation accompanied by
  voltage-dependent oscillations and characteristic algebraic decay.
\end{abstract}
\maketitle
\section{Introduction}

The standard set-up for a quantum dot consists of a small quantum system
described by a finite-dimensional Hilbert space which is coupled to several
infinitely large reservoirs via energy and/or particle exchange. A difference
in the chemical potentials of the reservoirs will generically lead to particle
transport and thus a finite current through the dot.  Here we will study the
arguably simplest but non-trivial quantum dot system, namely the interacting
resonant level model (IRLM). It is given by a local level coupled to two leads
of non-interacting spinless fermions. The fermions can hop on and off the
level. In addition, there is a Coulomb interaction between the level and the
reservoirs (see Fig.~\ref{fig:fig1}). The IRLM constitutes the minimal model 
for a quantum dot dominated by charge fluctuations as spin degrees of freedom 
are not taken into account.

Originally the (one-lead) IRLM was introduced independently by Wiegmann and
Finkelstein~\cite{VigmanFinkelstein78} as well as
Schlottmann~\cite{Schlottmann} to study the anisotropic Kondo model. They
generalized earlier works by Anderson et al.~\cite{Anderson} at the Toulouse
point~\cite{Toulouse}, where the Coulomb interaction between the level and the
reservoir vanishes. In particular, in
Refs.~\onlinecite{VigmanFinkelstein78,Schlottmann} it was shown that the IRLM
and the anisotropic Kondo model possess the same partition function in the
so-called long-time approximation and thus share the same universal
low-temperature characteristics. Equilibrium properties like the static and
dynamic susceptibilities and the relaxation rate of the IRLM have been
intensively studied in the early 1980's using the Bethe
Ansatz~\cite{FilyovWiegmann80} as well as renormalization group (RG)
techniques~\cite{Schlottmann82}. The equivalence between the IRLM and the
anisotropic Kondo model can be shown by bosonization and refermionization of
the latter model~\cite{GNT}.

Recently the interest in the IRLM has been revived as a minimal model
to describe non-equilibrium transport through quantum
dots. Initialized by the work of Mehta and Andrei~\cite{MehtaAndrei06}
the model has been investigated using the Hershfield
$Y$-operator~\cite{Doyon}, Keldysh perturbation~\cite{Golub} and
scattering theory~\cite{NH,NIH}, field theory
approaches~\cite{BS,BSS,BBSS}, the numerical renormalization group
method (NRG)~\cite{BVZ} and the time-dependent density matrix
renormalization group technique (TD-DMRG)~\cite{BSS,BBSS}. Most of
these studies were performed at the special point of particle-hole and
left-right symmetry. The quantity of main interest has been the
steady-state current through the resonant level. The main conclusions
were that (i) at sufficiently large bias voltages a negative
differential conductance appears, and that (ii) in the scaling limit,
where all bare energy scales are much smaller than the bandwidth of
the leads, the current decreases as a power law in the applied
voltage. However, only at the self-dual point~\cite{BSS} it has been
possible to derive closed analytic expressions for the current as a
function of the applied voltage.

Recently, perturbative RG techniques in
non-equilibrium~\cite{BVZ,karrasch10,SA} have been applied to obtain
more insight into the physics of the IRLM at finite bias.  In
Ref.~\onlinecite{BVZ} a poor man scaling analysis has been performed
up to next-to-leading order providing power-law exponents up to second
order in the Coulomb interaction.  In a non-equilibrium situation, the
RG flow was cut off heuristically by the voltage which induced an
emergent power-law behavior of the current as a function of the
voltage.  Subsequently, this analysis has been put on a more firm
basis by the application of recently developed RG methods in
non-equilibrium, the functional RG \cite{jakobs07} and the real-time
RG method \cite{S}, see Ref.~\onlinecite{review} for a recent
review. A short summary of the main results of the two methods for the
IRLM has been presented in Ref.~\onlinecite{SA}, where a leading order
expansion has been performed giving rise to power-law exponents linear
in the Coulomb interaction. The results were compared to numerically
exact NRG and DMRG methods and a good agreement has been observed for
moderate Coulomb interactions.  In particular, the conclusion was
drawn that power-law behavior does not take place in the generic case
of asymmetric Coulomb interactions between the dot and the left and
right reservoir. In addition, the scaling behavior at resonances away
from the particle-hole symmetric point has been
reexamined~\cite{Doyon}.  Details of the functional RG method have
been presented in Ref.~\onlinecite{karrasch10}.

In this paper we will present an extended version of
Ref.~\onlinecite{SA} concerning the real-time RG method, supplemented
by a generic treatment of strong charge fluctuations, a
next-to-leading order analysis in the Coulomb interaction and a new
result concerning power laws as function of the level position away
from the particle-hole symmetric point.  The real-time renormalization
group in frequency space (RTRG-FS)~\cite{S} has recently been
introduced in the theory of dissipative quantum systems.  It provides
a powerful tool in the description of non-equilibrium transport, in
particular the relaxation and decoherence rates naturally arise within
the proposed formalism. Previous applications to the Kondo
model~\cite{RTRGKondo,PSS} in the weak coupling regime are here
generalized to include charge fluctuations in strong coupling. To this
end we develop RG equations, where we expand all quantities around
zero Matsubara frequency, in contrast to previous treatments
\cite{RTRGKondo,PSS}, where a systematic expansion around the poor man
scaling solution has been performed. The RG equations are set up in a
generic form, which can also be used for other models with strong
charge fluctuations.  In particular, for the IRLM we demonstrate that
this scheme allows the study of observables close to resonances where
the tunneling rate is the only relevant energy scale quantifying
charge fluctuations. Furthermore, we extend the analysis in
Ref.~\onlinecite{SA} by including subleading terms, which gives the
exponents of power laws consistently up to second order in the Coulomb
interaction. A corresponding comparison of the power-law exponent with
NRG results for the charge susceptibility in equilibrium at the
particle-hole symmetric point yields excellent agreement. We present
approximate analytical solutions which are confirmed by numerically
integrating the corresponding full RG equations and which describe the
steady state as well as the quench dynamics for arbitrary system
parameters.  Thereby various microscopic cutoff scales of the RG flow
can be identified, which is essential for the precise determination of
the scaling behavior of observables. In particular, we derive closed
analytic expressions for the charge susceptibility, the steady-state
current and the differential conductance. We find (i) a negative
differential conductance for arbitrary system parameters, (ii) that
for asymmetric Coulomb interactions the current does in general not
follow a power law as a function of the bias and is recovered only in
the limit of extremely large voltages, (iii) that at resonance,
i.e. when the level position is aligned with one of the Fermi levels
in the leads, the current does not follow a power law even in the
symmetric model, and (iv) that the current or the linear conductance
reveals a power law as function of the level position in the generic
case, i.e. even for asymmetric Coulomb interactions and/or asymmetric
tunneling couplings. The latter result was not reported in
Ref.~\onlinecite{SA}.

In addition, we use the analytical solution of the RG equations to study the
quench dynamics in the IRLM, where we assume the couplings to the leads to be
switched on suddenly. We derive closed integral representations for the
resulting time evolution of the dot occupation and the current. The most
notable characteristics of the time evolution of both observables are:  (i)
the relaxation towards the stationary values is governed by two different
decay rates describing the charge relaxation on the level and its broadening
induced by the coupling to the leads, respectively, (ii) the voltage
appears as an important energy scale for the dynamics setting the frequency of
an oscillatory behavior, and (iii) the exponential decay is accompanied
by an algebraic behavior with an interaction-dependent
exponent. Similar results have been obtained recently for the dynamics
of the non-equilibrium Kondo model~\cite{PSS}, showing that these features are
generic.

The paper is organized as follows: In Section~\ref{sec:model} we introduce the
IRLM and discuss its description in Liouville space. In Section~\ref{sec:RG}
we summarize and solve the RG equations. In Section~\ref{sec:res} we present
the results for steady-state quantities as well as for the time evolution.
Here we also provide a simple derivation of the appearance of the negative
differential conductance.  Technical details together with the generic
derivation of non-equilibrium RG equations in the regime of strong charge
fluctuations are reported in the Appendix.

\section{Model}\label{sec:model}

\begin{figure}[t]
\includegraphics[width=80mm]{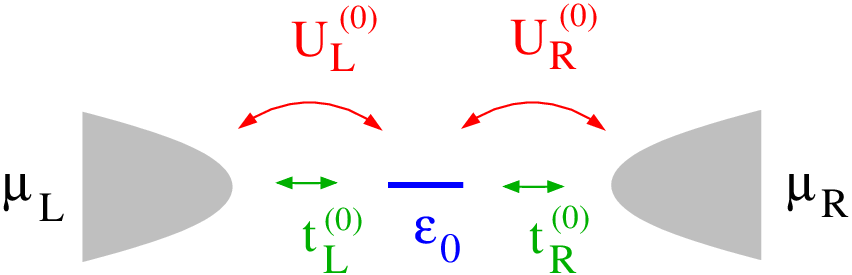}
\caption{(Color online) Sketch of the interacting resonant level
  model. A local level with energy $\epsilon_0$ is coupled via hoppings
  $t^{(0)}_{L/R}$ and Coulomb couplings $U^{(0)}_{L/R}$ to two spinless
  fermionic reservoirs held at chemical potentials $\mu_{L/R}=\pm V/2$.}
\label{fig:fig1}
\end{figure}

The Hamiltonian of the interacting resonant level
model (IRLM) depicted in Fig.~\ref{fig:fig1} is given by 
\begin{equation}
H=H_{res}+H_D+V\;,
\label{eq:model} 
\end{equation}
where 
\begin{equation}
H_{res}=\sum_{k\alpha}(\epsilon_k+\mu_\alpha) a_{k\alpha}^\dagger a_{k\alpha}
\end{equation}
describes a set of semi-infinite fermionic reservoirs with chemical
potentials $\mu_\alpha$. In the case of two reservoirs $\alpha=L/R$,
we choose $\mu_{L/R}=\pm V/2$.  Standard
second-quantized notation is used, and the energies
$\epsilon_{k\alpha}$ are restricted to a finite band of width $D$. The
dot Hamiltonian reads
\begin{equation}
H_D=\epsilon_0 c^\dagger c\;,
\end{equation}
and the fermionic level is coupled to the reservoirs via 
\begin{eqnarray}
  V&=&\sum_{\alpha}\frac{t_\alpha^{(0)}}{\sqrt{\rho^{(0)}_{\alpha}}} 
  \sum_k \bigl(a_{k\alpha}^\dagger c+c^\dagger a_{k\alpha}\bigr)\nonumber\\
  &&+\left(c^\dagger c-\frac{1}{2}\right)
  \sum_{\alpha}\frac{U_\alpha^{(0)}}{\rho^{(0)}_{\alpha}} 
  \sum_{kk'}{:\!a_{k\alpha}^\dagger a_{k'\alpha}\!:}\;, 
\end{eqnarray}
where $:\ldots:$ denotes normal-ordering, and $t_{\alpha}^{(0)}$ and
$U_{\alpha}^{(0)}$ are real.  In the following we denote the bare
parameters by the super-script ${}^{(0)}$.  In the scaling limit the
details of the frequency dependence of the local density of states in
the reservoirs $\rho_{\alpha}(\omega)$ do not play a significant role
as long as it is sufficiently regular on the energy scale of the
applied voltage, which allows us to appoximate it by a constant.  Following
Ref.~\onlinecite{S}, we choose the Lorentzian form
\begin{equation}
\label{eq:dos}
\rho_\alpha(\omega)\,=\,\rho^{(0)}_\alpha\,{D^2\over D^2+\omega^2}\;.
\end{equation}
We stress that the hybridization as well as the Coulomb interactions
to the leads are allowed to be asymmetric, which corresponds to a
generic setting.  Furthermore, we do not restrict ourselves to the
particle-hole symmetric point given by $\epsilon_0=0$.

We define
$a^{(\dagger)}_{\alpha}(\omega)=\frac{1}{\sqrt{\rho^{(0)}_{\alpha}}}\sum_k
\delta(\omega-\epsilon_k+\mu_\alpha) a_{k\alpha}^{(\dagger)}$, and
introduce the vertices
\begin{equation}
g_{\eta\alpha}=t_{\alpha}^{(0)}\left\{\begin{array}{ll}
c& \textrm{for $\;\eta=+$}\\
c^{\dagger}& \textrm{for $\;\eta=-$}
\end{array}\right.
\label{eq:defg1}
\end{equation}
and
\begin{equation}
g_{\eta\alpha,\eta'\alpha'}=\delta_{\eta,-\eta'}\delta_{\alpha,\alpha'}
\eta \,U_{\alpha}^{(0)}(c^{\dagger}c-\frac{1}{2})\;. 
\label{eq:defg11}
\end{equation}
The different contributions to the Hamiltonian can then be rewritten as
\begin{equation}
H_{res}=\sum_{\alpha}\int d\omega (\omega+\mu_{\alpha})
a^\dagger_{\alpha}(\omega)a_{\alpha}(\omega)\;,
\end{equation}
where we measure the energy $\omega$ of the reservoir states relative
to the chemical potentials $\mu_\alpha$,
\begin{equation}
H_D=\sum_s E_s \mid \!s \,\rangle\langle s\!\mid\;,
\end{equation}
with $s=0,1$, and $E_0=0$, $E_1=\epsilon_0$ respectively, and
\begin{eqnarray}
\label{eq:interaction}
V&=&\eta \int d\omega a_1(\omega)g_1\nonumber\\
&&+\frac{1}{2}\int d\omega\int d\omega'
\eta\eta'{:\!a_{1'}(\omega')a_1(\omega)\!:} g_{11'}\;,
\end{eqnarray}
with the multi-index $1\equiv \eta\alpha$ containing $\eta=\pm$ for
creation and annihilation operators and the lead index $\alpha$. 
Repeated incides are assumed to be summed over.
We consider the case of zero temperature throughout the manuscript since
temperature is a rather trivial cutoff parameter for the RG flow,
which at will can be easily incorporated in the employed RG formalism.

\section{RG analysis}\label{sec:RG}

We will study the non-equilibrium properties of the IRLM using the
real-time renormalization group method in frequency space~\cite{S}
(RTRG-FS).  The formalism is based on a description of the system in
Liouville space. The density matrix of the full system, $\rho(t)$, is
given by the solution of the von Neumann equation
\begin{eqnarray}
\rho(t)&=&e^{-i H(t-t_0)}\,\rho(t_0)\,e^{i H(t-t_0)}\nonumber\\
&=&e^{-i L(t-t_0)}\,\rho(t_0)\;,
\label{rhoformalL}
\end{eqnarray}
where $L=[H,.]$ is the Liouvillian acting on usual operators in
Hilbert space via the commutator.  Initially, we assume that the
density matrix is a product of an arbitrary dot part $\rho_D(t_0)$ and
grandcanonical distributions for the reservoirs,
\begin{equation}
  \rho(t_0)=\rho_D(t_0)\,\prod_{\alpha}\rho_{res}^\alpha\;.
  \label{eq:rhoinitial}
\end{equation}
The object of main interest is the reduced density matrix of the dot,
which is obtained by tracing out the reservoir degrees of freedom
\begin{equation}
  \rho_D(t)=\mbox{Tr}_{res}\,\rho(t)=
  \mbox{Tr}_{res}\,e^{-i L(t-t_0)}\,\rho_D(t_0)\,
\prod_{\alpha}\rho_{res}^\alpha\;,
\end{equation}
and its Laplace transform ($\mathrm{Im}\,z>0$)
\begin{equation}
  \tilde{\rho}_D(z)=\int_{t_0}^\infty dt\,e^{i z(t-t_0)}\,\rho_D(t)
  =\mbox{Tr}_{res}\,\frac{i}{z-L}\,\rho(t_0)\;.
\label{laplace}
\end{equation}
Here the Liouvillian admits the same decomposition as \eqref{eq:model}, i.e.
$L=L_{res}+L_D^{(0)}+L_V$ with $L_{res}=[H_{res},.]$, $L_D^{(0)}=[H_D,.]$, and
$L_V=[V,.]$.  Using the RTRG-FS we will derive the effective Liouvillian of
the quantum dot $L_D^{eff}(z)$ from which the reduced density matrix can be
calculated via
\begin{equation}
  \tilde{\rho}_D(z)={i\over z-L_D^{eff}(z)}\,\rho_D(t_0)\;.
  \label{eq:rhoDLeff}
\end{equation}
The stationary reduced density matrix is obtained as
\begin{equation}
  \rho^{st}_D=\lim_{t\rightarrow\infty} \rho_D(t)=
  \lim_{z\rightarrow i0^+}\frac{z}{z-L_D^{eff}(z)}\,\rho_D(t_0)\;.
  \label{eq:pstdef}
\end{equation}
The existence of a stationary density matrix was proven in Ref.~\onlinecite{S}
using the RTRG-FS as well as for the Kondo model in
Ref.~\onlinecite{DoyonAndrei06} using non-equilibrium perturbation theory to
all orders.  The matrix elements of the effective Liouvillian involve the
rates for the processes between the two eigenstates of the dot, leading to
poles of the resolvent \eqref{eq:rhoDLeff} at $z^1_{\mathrm{p}}=-i\Gamma_1$
and $z^\pm_{\mathrm{p}}=\pm \tilde{\epsilon}-i\Gamma_2$, where $\Gamma_1$
corresponds to the charge relaxation rate, $\Gamma_2$ describes half of the
broadening of the local level, and $\tilde{\epsilon}$ is the renormalized level
position.

The calculation of the current follows along the same lines. The
operator for the particle current flowing from reservoir $\gamma$ to
the dot is defined as $I^\gamma=-dN^\gamma/dt=-i[H,N^\gamma]$, where
$N^\gamma$ denotes the corresponding particle number operator in lead
$\gamma$. The current in lead $\gamma$ then reads $\langle
I^\gamma\rangle(t)=\mbox{Tr}_{D}\mbox{Tr}_{res}\,I^\gamma\,\rho(t)$.
Tracing out the reservoir degrees of freedom it can be written as
\begin{equation}
  \langle I^\gamma\rangle(z)=
  -i\,\mbox{Tr}_{D}\,\Sigma_\gamma(z)\,\tilde{\rho}_D(z)
  \label{eq:Current}
\end{equation}
in Laplace space, where $\Sigma_\gamma(z)$ denotes the current kernel to be
derived below. The stationary current is given by $\langle
I^\gamma\rangle^{st}=-i\,\mbox{Tr}_{D}\,\Sigma_\gamma(i0^+)\,\rho_D^{st}$.

Through $L_D^{eff}(z)$, the RTRG-FS method provides direct access to the
microscopic cutoff scales.  By systematically integrating out the energy
scales of the reservoirs step by step, a formally exact RG equation can be
derived for $L_D^{eff}(z)$ as a function of a flow parameter $\Lambda$, where
all reservoir energy scales beyond $\Lambda$ are included.  This RG equation
is coupled to other RG equations for the couplings. Similar schemes can be
developed for the calculation of the transport current and correlation
functions \cite{S,RTRGKondo}. All RG equations involve resolvents similar to
the one occurring in \eqref{eq:rhoDLeff}, where $z$ is shifted by the physical
energy scales like the reservoirs electrochemical potentials.  The cutoff scale
is given by the distance to resonances, being replaced by the corresponding
rate at resonance.  The microscopic inclusion of decay rates as cutoff scales
into non-equilibrium RG methods was also achieved within flow equation methods
\cite{Kehrein05}.

\subsection{Parametrization and initial conditions}

In Liouville space, defined by the basis $(00\;11\;10\;01)$, the bare
Liouvillian is given by $L_D^{(0)}=[H_D,\cdot]$, and the bare vertices are
\begin{equation}
\label{eq:single_vertices}
G_1^{p(0)}=\sigma^p\left\{\begin{array}{ll}
g_1 \cdot & \textrm{for $\;p=+$}\\
- \cdot g_1 & \textrm{for $\;p=-$}
\end{array}\right.
\end{equation}
and
\begin{equation}
\label{eq:double_vertices}
G_{11'}^{pp'(0)}=\delta_{pp'}\left\{\begin{array}{ll}
g_{11'} \cdot & \textrm{for $\;p=+$}\\
- \cdot g_{11'} & \textrm{for $\;p=-$}
\end{array}\right.\;,
\end{equation}
where $\sigma^+=\mathbb{I}$ and 
 \begin{equation}
\sigma^-= \left( \begin{array}{cccc}
 1&0&0&0 \\
 0&1 &0&0 \\
 0&0&-1&0\\
0&0  &0&-1 \end{array} \right) \;.
\end{equation}
The bare current vertex reads
$(I^{\gamma})^{p(0)}_1=-\frac{1}{2}\eta \delta_{\alpha\gamma}p\,
G_1^{p(0)}$. 

For the vertices the following notations are introduced:
\begin{eqnarray}
\label{eq:single_vertices_bt}
\bar{G}_1^{(0)}&=&\sum_pG_1^{p(0)} \quad,\quad
\tilde{G}_1^{(0)}=\sum_pp\,G_1^{p(0)}\quad,\\
\label{eq:double_vertices_bt}
\bar{G}_{11'}^{(0)}&=&\sum_pG_{11'}^{pp(0)} \quad,\quad
\tilde{G}_{11'}^{(0)}=\sum_pp\,G_{11'}^{pp(0)}\quad,
\end{eqnarray}
together with $\bar{I}_1^{\gamma(0)}=\sum_p(I^{\gamma})^{p(0)}_1$.  We
note that \eqref{eq:single_vertices_bt} and
\eqref{eq:double_vertices_bt} are related to the commutators and
anticommutators of \eqref{eq:defg1} and \eqref{eq:defg11},
respectively.  In matrix notation, the bare Liouvillian and the bare
vertices are then given by
\begin{eqnarray}
\nonumber
L_D^{(0)}&=& 
\epsilon_0\left( \begin{array}{cccc}
0&0&0&0 \\
0&0&0&0 \\
0&0&1&0\\
0&0&0&-1 \end{array} \right)\;,\\
\nonumber
\bar{G}_{+\alpha}^{(0)}&=&
t_\alpha^{(0)}\left( \begin{array}{cccc}
 0&0&1&0 \\
 0&0&-1&0 \\
 0&0&0&0\\
 1&1&0& 0\end{array} \right)\;,\\
\nonumber
\bar{G}_{-\alpha}^{(0)}&=&
t_\alpha^{(0)}\left( \begin{array}{cccc}
 0&0&0&-1 \\
 0&0&0&1 \\
 1&1&0&0\\
 0&0&0&0\end{array} \right)\;,\\
\nonumber
\tilde{G}_{+\alpha}^{(0)}&=&
t_\alpha^{(0)}\left( \begin{array}{cccc}
 0&0&1&0 \\
 0&0&1&0 \\
 0&0&0&0\\
-1&1&0& 0\end{array} \right)\;,\\
\nonumber
\tilde{G}_{-\alpha}^{(0)}&=&
t_\alpha^{(0)}\left( \begin{array}{cccc}
 0&0&0&1 \\
 0&0&0&1 \\
 1&-1&0&0\\
 0&0&0&0\end{array} \right)\;,\\
\nonumber
\bar{G}_{+\alpha,-\alpha}^{(0)}&=& 
U_\alpha^{(0)}\left( \begin{array}{cccc}
 0&0&0&0 \\
 0&0&0&0 \\
 0&0&1&0\\
 0&0&0&-1\end{array} \right)\;,\\
\label{eq:bare}
\tilde{G}_{+\alpha,-\alpha}^{(0)}&=& 
U_\alpha^{(0)}\left( \begin{array}{cccc}
 -1&0&0&0 \\
 0&1&0&0 \\
0&0 &0& 0\\
 0&0&0&0 \end{array} \right)\;,
\end{eqnarray}
with $\bar{G}_{-\alpha,+\alpha}^{(0)}=-\bar{G}_{+\alpha,-\alpha}^{(0)}$
and $\tilde{G}_{-\alpha,+\alpha}^{(0)}=-\tilde{G}_{+\alpha,-\alpha}^{(0)}$.
For the current vertex we obtain
\begin{eqnarray}
\label{eq:bare_ivert}
{\rm Tr}_D \, \bar{I}_{+\alpha}^{\gamma(0)}&=&
-\delta_{\alpha\gamma}t_{\alpha}^{(0)}(0\;\;0\;\;1\;\;0)
\nonumber\\
{\rm Tr}_D  \,\bar{I}_{-\alpha}^{\gamma(0)}&=&
\delta_{\alpha\gamma}t_{\alpha}^{(0)}(0\;\;0\;\;0\;\;1)\;.
\end{eqnarray}

Within the RG treatment, the Liouvillian $L_D(z)$ and the vertices
$\bar{G}_1(z;\omega_1)$, $\bar{G}_{11'}(z;\omega_1,\omega_1^\prime)$
and $\bar{I}_1^\gamma(z;\omega_1)$ are effective quantities, which
obtain an additional dependence on the Laplace variable $z$ and depend
on frequency variables $\omega_1$ and $\omega_1^\prime$ (the vertices
$\tilde{G}_1^{(0)}$ and $\tilde{G}_{11'}^{(0)}$ are only needed for
the initial setup of the RG flow). In addition the current kernel
$\Sigma_\gamma(z)$ is generated.  As shown in App.~\ref{app:flow}, the
dependence of the vertices on the frequencies $\omega_1$ and
$\omega_1^\prime$ can be treated in leading order by expanding around
$\omega_1=\omega_1^\prime=0$. Therefore, we omit it in the following
and, furthermore, replace $z$ by its real part
$E\equiv\text{Re}\{z\}$. The full $z$-dependence can be recovered
finally by analytic continuation, which will be done in
Sec.~\ref{sec:time} where we study the time evolution.

Following Ref.~\onlinecite{S}, the parametrization of the renormalized
quantities follows from charge conservation and the following symmetry
properties
\begin{eqnarray}
\nonumber
{\rm Tr}_D L_D(E)&=&
{\rm Tr}_D \bar{G}_1(E)=
{\rm  Tr}_D \bar{G}_{11'}(E)=0\;,\\
\nonumber
L_D(E)^c &=& -L_D(-E)\;,\quad\Sigma_\gamma(E)^c=-\Sigma_\gamma(-E)\;,\\
\nonumber
\bar{G}_1(E)^c &=& -\sigma^-\bar{G}_{\bar{1}}(-E)\;,\quad
\bar{G}_{11'}(E)^c = \bar{G}_{\bar{1}\bar{1}'}(-E)\;,\\
\nonumber
\bar{I}_1^\gamma(E)^c &=& -\sigma^- \bar{I}_{\bar{1}}(-E)^\gamma\;,
\end{eqnarray}
where $(A^c)_{s_1s_{1'},s_2s_{2'}}=A^*_{s_{1'}s_1,s_{2'}s_2}$ and
$\bar{1}\equiv -\eta\alpha$.

As a consequence, the renormalized Liouvillian can be written as
\begin{equation}
\label{eq:L_form}
L_D(E)= \left( \begin{array}{cccc}
 -i \Gamma_+(E)&i \Gamma_-(E)&0&0 \\
 i \Gamma_+(E)&-i \Gamma_-(E) &0&0 \\
 0&0&\epsilon(E)&0\\
0&0  &0&-\epsilon(-E)^* \end{array} \right)\;,
\end{equation}
with $\Gamma_{\pm}(E)=\Gamma_{\pm}(-E)^*$. The renormalized vertices are given
by
\begin{eqnarray}
\label{eq:vert1}
\!\!\!\!\!\!\!\!\!\!\!\!\!\bar{G}_{+\alpha}(E)&= &\left( \begin{array}{cccc}
 0&0&t_\alpha(E)&0 \\
 0&0 &-t_\alpha(E)&0 \\
 0&0&0&0\\
t_\alpha^2(E)&t_\alpha^3(E)  &0& 0\end{array} \right)\;,\nonumber\\
\!\!\!\!\!\!\!\!\!\!\!\!\!\bar{G}_{-\alpha}(E)&= &\left( \begin{array}{cccc}
 0&0&0&-t_\alpha(-E)^* \\
 0&0 &0&t_\alpha(-E)^* \\
t_\alpha^2(-E)^*&t_\alpha^3 (-E)^* &0& 0\\
 0&0&0&0\end{array} \right)\;,
\end{eqnarray}
and
\begin{equation}
\label{eq:vert2}
\bar{G}_{+\alpha,-\alpha}(E)= \left( \begin{array}{cccc}
 0&0&0&0 \\
 0&0 &0&0 \\
0&0 &U_{\alpha}(E)& 0\\
 0&0&0&-U_{\alpha}(-E)^* \end{array} \right)\;,
\end{equation}
with $\bar{G}_{-\alpha,+\alpha}(E)=-\bar{G}_{+\alpha,-\alpha}(E)$.
This form of $\bar{G}_{11'}(E)$ holds only in leading order, as
higher-order RG contributions generate non-zero elements in the upper
left $2 \times 2$ block (see Sec.~\ref{sec:flow}), while the form (\ref{eq:L_form}) 
and (\ref{eq:vert1}) are retained to all orders.
For the renormalized current vertex we obtain the parametrization 
\begin{eqnarray}
\label{eq:ivert}
{\rm Tr}_D \, \bar{I}_{+\alpha}^{\gamma}(E)&=&
-t_{\alpha}^{\gamma}(E)\;(0\;\;0\;\;1\;\;0)
\nonumber\\
{\rm Tr}_D  \,\bar{I}_{-\alpha}^{\gamma}(E)&=&
t_{\alpha}^{\gamma}(-E)^*\;(0\;\;0\;\;0\;\;1)\;,
\end{eqnarray}
as well as for the corresponding current kernel generated by the RG flow
\begin{eqnarray}
\label{eq:ikernel}
{\rm Tr}_D \,\Sigma_{\gamma}(E)=i(\Gamma^1_{\gamma}(E)\;\;
\Gamma^2_{\gamma}(E)\;\;0\;\;0)\;,
\end{eqnarray}
with $\Gamma^i_{\gamma}(E)=\Gamma^i_{\gamma}(-E)^*$.

The bare values, which serve as initial conditions for the RG
equations, read $\epsilon(E)=\epsilon_0$, $\Gamma_{\pm}(E)=0$,
$t_\alpha(E)=t_\alpha^2(E)=t_\alpha^3(E)=t_{\alpha}^{(0)}$,
$U_{\alpha}(E)=U_{\alpha}^{(0)}$, $\Gamma^i_{\gamma}(E)=0$, and
$t_{\alpha}^{\gamma}(E)=\delta_{\alpha\gamma}t_{\alpha}^{(0)}$.

\subsection{Flow equations}\label{sec:flow}

In this section we summarize the RG equations for the renormalized quantities
as introduced in the previous section, a detailed derivation is given in
App.~\ref{app:flow}.

The diagrams taken into account are shown in Fig.~\ref{fig:fig3}. We consider
contributions to the flow of $L_D$, $\bar{G}_1$ and $\bar{G}_{11'}$ to lowest
order in $\Gamma\sim t^2$ to describe the scaling limit and to leading and
next-to-leading order in $U_{\alpha}$ to obtain exponents up to order
$O(U_{\alpha}^2)$.  Terms of order $\sim \Gamma U_{\alpha}$ for
$\bar{G}_{11'}$ are neglected. These would generate nonzero elements in the
upper left $2\times2$ block of \eqref{eq:vert2}.  For the Liouvillian and the
vertices the full $E$-dependence crucial for the time evolution is taken into
account.

Basing on the parametrization of the Liouvillian, the current kernel,
and the vertices, we introduce the following definitions
\begin{eqnarray}
\nonumber
Z(E) &=& \left(1-{d\over dE}\epsilon(E)\right)^{-1}\;,\\
\nonumber
\tilde{\Gamma}_\alpha(E) &=& 2\pi Z(E+\mu_\alpha)t_\alpha(E)^2\;,
\phantom{\frac{1}{2}}\\
\nonumber
\Gamma_\alpha(E) &=& \Gamma_\alpha^1(E) - \Gamma_\alpha^2(E) \;,
\phantom{\frac{1}{2}}\\
\nonumber
\Gamma^\prime_\alpha(E) &=& {1\over 2}\left(\Gamma_\alpha^1(E) + 
\Gamma_\alpha^2(E)\right) \;,\phantom{\frac{1}{2}}\\
\nonumber
\Gamma(E) &=& \sum_\alpha \Gamma_\alpha(E) \;,\quad \Gamma^\prime(E) = 
\sum_\alpha \Gamma^\prime_\alpha(E)\;,\phantom{\frac{1}{2}}\\
\nonumber
\chi(E) &=& Z(E)\left(E-\epsilon(E)\right)\;,\phantom{\frac{1}{2}}\\
\nonumber
\chi'(E) &=& \chi(E)-2i\gamma_0\Lambda \ln{2\Lambda-i\chi(E)\over 
\Lambda-i\chi(E)}\;,\\
\label{eq:definitions}&&
\end{eqnarray} 
where $\gamma_0=\sum_{\alpha}(U_{\alpha}^{(0)})^2$, and $\Lambda$ is a
high-energy cutoff which cuts off the Matsubara frequencies of the Fermi
functions of the reservoirs.  Under the RG the cutoff parameter $\Lambda$
flows from the initial value $\Lambda_0$ to zero. The initial cutoff is
related to the physical reservoir band width $D$ by
\eqref{eq:D_initial_cutoff}, see App.~\ref{app:flow}.  As shown in
App.~\ref{app:flow}, the flow equations for the effective model parameters
read
\begin{eqnarray}
\label{eq:rg_gamma_tilde}
\frac{d}{d\Lambda}\tilde{\Gamma}_{\alpha}(E)&=&\\
\nonumber
&& \hspace{-1.5cm}  -\left( \frac{2(U_{\alpha}^{(0)}-\gamma_0)}
{\Lambda-i\chi'(E+\mu_{\alpha})}+
\frac{\gamma_0}{\Lambda-i\chi'(E+\mu_{\alpha})/2}\right)
\tilde{\Gamma}_{\alpha}(E)\;,\\
\label{eq:rg_gamma}
\frac{d}{d\Lambda}\Gamma_{\alpha}(E)&=&\\
\nonumber 
&&\hspace{-1.5cm} -\frac{U_{\alpha}^{(0)}}{\Lambda-i\chi'(E+\mu_{\alpha})}
\tilde{\Gamma}_{\alpha}(E)+(E\to-E)^*\;,\\
\label{eq:rg_gamma_prime}
\frac{d}{d\Lambda}\Gamma'_{\alpha}(E)&=&\\
\nonumber 
&&\hspace{-1.5cm}  \frac{i}{2\pi} \frac{1}{\Lambda-i\chi'(E+\mu_{\alpha})}
\tilde{\Gamma}_{\alpha}(E)+(E\to-E)^*\;,\\
\label{eq:rg_chi}
\frac{d}{d\Lambda}\chi'(E)&=& \\
\nonumber
&&\hspace{-1.5cm} -i\sum_{\alpha} \frac{U_{\alpha}^{(0)}}
{\Lambda+\Gamma(E-\mu_{\alpha})-i(E-\mu_{\alpha})}
\tilde{\Gamma}_{\alpha}(E-\mu_{\alpha})\;.
\end{eqnarray}
The remaining parameters of the Liouvillian and the vertices are given by
\begin{eqnarray}
\!\!\!\!\!\!\!\!\Gamma_\pm(E) &=& {1\over 2}\Gamma(E)\pm \Gamma^\prime(E)
= \pm \sum_\alpha \Gamma_\alpha^{1/2}(E)\;,
\label{eq:solution_gamma_pm}\\
\label{eq:solution_t_23}
\!\!\!\!\!\!\!\!t_{2/3}^\alpha(E) &=& 
t_\alpha(E)(1\pm i\pi U_\alpha^{(0)})\;,\\
\label{eq:solution_t_current}
\!\!\!\!\!\!\!\!t_\alpha^\gamma(E) &=& \delta_{\alpha\gamma}t_\alpha(E)\;,\\
\label{eq:solution_U}
\!\!\!\!\!\!\!\!Z(E)\,U_\alpha(E) &=& U_\alpha^{(0)}\;.
\end{eqnarray}
As a consequence, it turns out that $Z(E)U_{\alpha}(E)$ is unrenormalized up
to the second order in the interaction, in agreement with previous results
\cite{BVZ}.

The initial conditions for the RG equations are
$\tilde{\Gamma}_{\alpha}(E)=\Gamma_{\alpha}(E)=\Gamma_{\alpha}^{(0)}=2\pi
(t_{\alpha}^{(0)})^2$, $\Gamma'_{\alpha}(E)=0$ and
$\chi'(E)=E-\epsilon_0+\frac{i}{2}\Gamma^{(0)}$, where
$\Gamma^{(0)}=\sum_{\alpha}\Gamma^{(0)}_{\alpha}$.  For the numerical solution
of (\ref{eq:rg_gamma_tilde})-(\ref{eq:rg_chi}) a discretization in $E$ is
required, the involved numerical effort is however limited due to the fast
convergence.

The RG equations (\ref{eq:rg_gamma_tilde})-(\ref{eq:rg_chi}) reduce to
poor man scaling equations for large $\Lambda$, where all resolvents
can be replaced by $1/\Lambda$. In this case similar power laws are
obtained for the stationary current as in Ref.~\onlinecite{BVZ},
provided that the cutoff parameter is intuitively inserted by hand. In
contrast, the RG equations derived in this paper reveal
microscopically the various cutoff parameters. As can be seen from
(\ref{eq:rg_gamma_tilde})-(\ref{eq:rg_gamma_prime}), all rates are cut
off by the distance to resonances, given by $\chi'(E+\mu_\alpha)$. On
the other hand, we see from (\ref{eq:rg_chi}) that the renormalization
of the level broadening, which is contained in the imaginary part of
$\chi'(E)$, is cut off by $|E-\mu_\alpha-i\Gamma(E-\mu_\alpha)|$.  The
RG equations presented here go beyond all previous RG analysis for the
IRLM.  Whereas Ref.~\onlinecite{BVZ} provided a consistent poor man
scaling analysis without a microscopic derivation of the cutoff
scales, Refs.~\onlinecite{karrasch10,SA} showed results from a full
microscopic non-equilibrium RG analysis, but only in leading order in
$U_\alpha^{(0)}$ for the exponent.

\subsection{Analytical solution}

Within the RTRG-FS approach, the coupled differential equations for the flow
of the effective system parameters as a function of the infrared cutoff
$\Lambda$ can be solved analytically. The approximate solutions are confirmed
by numerically integrating the corresponding full RG equations
(\ref{eq:rg_gamma_tilde})-(\ref{eq:rg_chi}).

The poor man scaling version of (\ref{eq:rg_gamma_tilde}), i.e. where the
resolvents are replaced by $1/\Lambda$, gives the power-law solution
\begin{equation}
\label{eq:pms_gamma_tilde}
\tilde{\Gamma}_\alpha\rightarrow\Gamma_\alpha^{(0)}
\left(\Lambda_0/\Lambda\right)^{g_\alpha}\quad,
\end{equation}
with the exponent
\begin{equation}
\label{eq:exponent}
g_\alpha = 2U_\alpha^{(0)}-\gamma_0 = 
2U_\alpha^{(0)}-\sum_\beta (U_\beta^{(0)})^2\;.
\end{equation}
According to (\ref{eq:rg_gamma_tilde}) this power law is cut off by
$\chi'(E+\mu_\alpha)$.  Therefore, the leading order solution is given by
\begin{eqnarray}
\label{eq:gamt}
\tilde{\Gamma}_{\alpha}(E)&\simeq&  
\Gamma^{(0)}_{\alpha} \left( \frac{\Lambda_0}
{\Lambda-i\chi'(E+\mu_{\alpha})}\right)^{g_{\alpha}}\;,
\end{eqnarray}
where the exponent is consitently calculated up to $O(U^2)$.

Since, for small $U_\alpha^{(0)}\ll 1$, the power laws lead only to a weak
variation, we can use the poor man scaling solution (\ref{eq:pms_gamma_tilde})
for $\tilde{\Gamma}_\alpha$ in the other RG equations
(\ref{eq:rg_gamma})-(\ref{eq:rg_chi}), and read off the cutoff scale by the
remaining resolvents in these equations. This gives the following leading
order solution
\begin{eqnarray}
\label{eq:solution_gamma}
\Gamma_{\alpha}(E)&\simeq&
\frac{1}{2}(\tilde{\Gamma}_{\alpha}(E)+\tilde{\Gamma}_{\alpha}(-E)^*)\;,\\
\label{eq:solution_gamma_prime}
\Gamma'_{\alpha}(E)&\simeq&-\frac{i}{4\pi U_{\alpha}^{(0)}}
(\tilde{\Gamma}_{\alpha}(E)-\tilde{\Gamma}_{\alpha}(-E)^*)\;,\\
\label{eq:solution_chi}
\chi'(E) &\simeq& E-\epsilon_0+\frac{i}{2}\Gamma_\epsilon(E)\;,
\end{eqnarray}
with the renormalized level broadening
\begin{equation}
\label{eq:solution_gamma_epsilon}
  \Gamma_\epsilon(E)=\sum_{\alpha}\Gamma_\alpha^{(0)}
  \left(\frac{\Lambda_0}{\Lambda+\Gamma(E-\mu_\alpha)-i(E-\mu_\alpha)}
  \right)^{g_\alpha}\;.
\end{equation}
We note the properties
\begin{eqnarray}
\label{eq:gamma_property}
\Gamma_\alpha(E)^* &=& \Gamma_\alpha(-E)\quad,\\
\label{eq:gamma_prime_property}
\Gamma^\prime_\alpha(E)^* &=& \Gamma^\prime_\alpha(-E)\quad.
\end{eqnarray}

In the limit $\Lambda\rightarrow 0$, we obtain 
\begin{eqnarray}
  \label{eq:gamma_epsilon_final}
\!\!\!\!\!\!\!\! \!\!\!\! \Gamma_\epsilon(E) &=& 
\sum_{\alpha}\Gamma_\alpha^{(0)}
  \left(\frac{\Lambda_0}{\Gamma(E-\mu_\alpha)-i(E-\mu_\alpha)}
  \right)^{g_\alpha}\;,\\
\!\!\!\!\!\!\!\!\!\!\!\!  \tilde{\Gamma}_\alpha(E) &=& \Gamma_\alpha^{(0)}
  \left(\frac{\Lambda_0}{\frac{1}{2}\Gamma_\epsilon(E+\mu_\alpha)
      -i(E+\mu_\alpha-\epsilon_0)}\right)^{g_\alpha}\;,
  \label{eq:gamma_tilde_final}
\end{eqnarray}
which, together with \eqref{eq:solution_gamma} gives a self-consistent set of
equations for the determination of $\Gamma_\epsilon(E)$ and $\Gamma(E)$. In
principle this set can be solved numerically but we will provide further
analytic evaluations in Section~\ref{sec:res}.

The reduced density matrix $\tilde{\rho}_D(E)=(p_0(E)\;\;p_1(E)\;\;0\;\;0)^T$
of the dot in Laplace space can be obtained from (\ref{eq:rhoDLeff}) and
(\ref{eq:L_form}), with $L_D^{eff}(E)\equiv L_D(E)|_{\Lambda=0}$. After a
straightforward algebra we obtain
\begin{equation}
\label{eq:p_01}
p_{0/1}(E) = {i\over E}p_{0/1}(t_0)
+{\Gamma(E)p_{0/1}(t_0)-\Gamma_\mp(E) \over E(E+i\Gamma(E))}\;,
\end{equation}
where $p_{0/1}(t_0)$ are the initial occupation probabilities for the dot and
$\Gamma_\pm(E)=\Gamma(E)/2\pm\Gamma'(E)$, according to
(\ref{eq:solution_gamma_pm}).

Finally, using (\ref{eq:Current}) and (\ref{eq:ikernel}), the current in
Laplace space is computed using the density matrix by
\begin{eqnarray}
\label{eq:curr_laplace}
\langle I_{\alpha} \rangle (E)&=&
-i {\rm Tr}_D \,\Sigma_{\alpha}(E)\tilde{\rho}_D(E)\nonumber\\
        &=&\Gamma^1_{\alpha}(E)p_0(E)+\Gamma^2_{\alpha}(E)p_1(E)\;,
\end{eqnarray}
where
$\Gamma^{1/2}_{\alpha}(E)=\Gamma'_{\alpha}(E)\pm\frac{1}{2}\Gamma_{\alpha}(E)$,
according to (\ref{eq:definitions}).

The stationary probabilities $p_{0/1}^{st}$ and the stationary current
$I^{st}_\alpha$ follow from $p_{0/1}^{st}=\lim_{E\rightarrow 0}(-i)E
\,p_{0/1}(E)$ and $I^{st}=\Gamma^1_{\alpha}p_0^{st}+\Gamma^2_\alpha p_1^{st}$,
with $\Gamma^i_\alpha\equiv\Gamma^i_\alpha(E=0)$. Using (\ref{eq:p_01}) this
gives
\begin{eqnarray}
\label{eq:p_stationary}
p_{0/1}^{st} &=& {1\over 2}\mp {\Gamma'\over\Gamma}\quad,\\
\label{eq:I_stationary}
I_\alpha^{st} &=& \Gamma_\alpha^\prime 
- {\Gamma'\over\Gamma}\Gamma_\alpha\quad,
\end{eqnarray}
where all rates are evaluated at $E=0$. As required, we obtain conservation of
probability $p_0^{st}+p_1^{st}=1$ as well as current conservation $\sum_\alpha
I^{st}_\alpha=0$.

\section{Results}\label{sec:res}
\subsection{Steady-state quantities}

The stationary state is obtained for $E=0$ from (\ref{eq:p_stationary}) and
(\ref{eq:I_stationary}). Inserting the solution \eqref{eq:solution_gamma} and
\eqref{eq:solution_gamma_prime} for $\Gamma_\alpha$ and $\Gamma'_\alpha$
together with the expression \eqref{eq:gamma_tilde_final} for
$\tilde{\Gamma}_\alpha$, we obtain
$\Gamma_\alpha=\text{Re}\tilde{\Gamma}_\alpha$ and $\Gamma'_\alpha={1\over
  2\pi U_\alpha^{(0)}}\text{Im}\tilde{\Gamma}_\alpha$, with
\begin{equation}
\tilde{\Gamma}_\alpha = \Gamma_\alpha^{(0)}\left({\Lambda_0\over
{1\over 2}\Gamma_\epsilon(\mu_\alpha)
-i(\mu_\alpha-\epsilon_0)}\right)^{g_\alpha}\;.
\end{equation}
Since the cutoff $\Gamma_\epsilon(\mu_\alpha)$ is only relevant for
$|\mu_\alpha-\epsilon_0|\sim O(\Gamma)$ and since $\Gamma_\epsilon(E)$ varies
only weakly as function of $E$, we can replace with good accuracy
$\Gamma_\epsilon(\mu_\alpha)\rightarrow\Gamma_\epsilon(\epsilon_0)$ in the
last equation. Furthermore, neglecting terms with higher powers in
$U_\alpha^{(0)}$, we find in leading order
\begin{eqnarray}
\label{eq:gamma_stat}
\Gamma_\alpha &\simeq& \Gamma_\alpha^{(0)}\left({\Lambda_0\over
|{1\over 2}\Gamma_\epsilon(\epsilon_0)-
i(\mu_\alpha-\epsilon_0)|}\right)^{g_\alpha}\;,\\
\label{eq:gamma_prime_stat}
\Gamma'_\alpha &\simeq& {1\over\pi}\Gamma_\alpha
\arctan{\mu_\alpha-\epsilon_0 \over \Gamma_\epsilon(\epsilon_0)/2}\;.
\end{eqnarray}
To determine the level broadening $\Gamma_\epsilon(\epsilon_0)$, we use
\eqref{eq:gamma_epsilon_final} and replace
$\Gamma(\epsilon_0-\mu_\alpha)\rightarrow\Gamma(0)\equiv\Gamma$ in this
equation by using the same arguments as above. In leading order in
$U_\alpha^{(0)}$ this gives
\begin{equation}
\label{eq:gamma_epsilon_stat}
\Gamma_\epsilon(\epsilon_0) \simeq\sum_\alpha\Gamma_\alpha^{(0)}\left({\Lambda_0\over
|\Gamma+i(\mu_\alpha-\epsilon_0)|}\right)^{g_\alpha}\;.
\end{equation}
Neglecting the factor ${1\over 2}$ for the cutoff parameter
$\Gamma_\epsilon(\epsilon_0)$ in \eqref{eq:gamma_stat}, the self-consistent
solution of \eqref{eq:gamma_epsilon_stat} and \eqref{eq:gamma_stat} is
approximately
\begin{equation}
\label{eq:gamma_vs_broadening}
\Gamma\,\simeq\,\Gamma_\epsilon(\epsilon_0)\;.
\end{equation}

Inserting \eqref{eq:gamma_prime_stat} into \eqref{eq:p_stationary} and
\eqref{eq:I_stationary}, and using \eqref{eq:gamma_vs_broadening}, we find for 
the stationary dot occupation $n^{st}=p_1^{st}$
and the stationary current $I_\alpha^{st}$
\begin{eqnarray}
\label{eq:n_general}
n^{st} &=& {1\over 2}+{1\over \pi}\sum_\alpha {\Gamma_\alpha\over\Gamma}
\arctan{\mu_\alpha-\epsilon_0\over\Gamma/2}    \\
\nonumber
I_\alpha^{st} &=& 
G_0\sum_{\beta\ne\alpha}{2\Gamma_\alpha\Gamma_\beta \over \Gamma}\\
\label{eq:curr_general}
&&\hspace{-0cm}\times\!\left(\arctan{\mu_\alpha-\epsilon_0\over\Gamma/2}-
\arctan{\mu_\beta-\epsilon_0\over\Gamma/2}\right)\;,
\end{eqnarray}
with $G_0={e^2\over h}={1\over 2\pi}$ in our units. As a consequence, we find
in leading order the same form as in the noninteracting case (where the result
is exact) with Lorentzian resonances for the differential conductance at
$\mu_\alpha=\epsilon_0$. However, the rates $\Gamma_\alpha$ entering these
equations are not the bare ones but are strongly renormalized by the
interaction.  According to \eqref{eq:gamma_stat} and
\eqref{eq:gamma_vs_broadening} they have to be determined from the
self-consistent equation
\begin{equation}
\label{eq:gamma_sc}
\Gamma_\alpha \simeq \Gamma_\alpha^{(0)}\left({\Lambda_0\over
|{1\over 2}\Gamma-i(\mu_\alpha-\epsilon_0)|}\right)^{g_\alpha}\;,
\end{equation}
with $\Gamma=\sum_\alpha\Gamma_\alpha$. This equation will be further analyzed 
in the next section. In particular, this renormalization is responsible for a 
negative differential conductance at large voltage.

For simplicity, we will restrict ourselves in the following mainly to the case
of two reservoirs $\alpha=L,R$ with $\mu_L=-\mu_R=V/2$. In this case the dot
occupation and the current $I^{st}\equiv I^{st}_L =-I^{st}_R$ read
\begin{equation}
\label{eq:n}
n^{st}=\frac{1}{2}+\frac{1}{\pi}
  \left(\frac{\Gamma_L}{\Gamma}\,
    \text{arctan}\frac{V/2 -\epsilon_0}{\Gamma/2}-
    \frac{\Gamma_R}{\Gamma}\,
    \text{arctan}\frac{V/2 + \epsilon_0}{\Gamma/2}\right)
\end{equation}
and
\begin{equation}
  \label{eq:curr}
  I^{st}=G_0\frac{2\Gamma_L\Gamma_R}{\Gamma}
  \left(\text{arctan}\frac{V/2-\epsilon_0}{\Gamma/2}+
    \text{arctan}\frac{V/2+ \epsilon_0}{\Gamma/2}\right)\;.
\end{equation}

\subsubsection{The rates $\Gamma_{\alpha}$}

As outlined above the rates $\Gamma_\alpha$ are determined by the 
self-consistent equation \eqref{eq:gamma_sc}. We define the cutoff scales
\begin{equation}
\label{eq:cutoff_alpha}
\Lambda_c^\alpha = \max\{|\mu_{\alpha}-\epsilon_0|,\frac{\Gamma}{2} \}\;.
\end{equation}
From \eqref{eq:gamma_sc} we see that $\Gamma_\alpha$ is renormalized by a
power law cut off by $\Lambda_c^\alpha$ 
\begin{equation}
\label{eq:gamma_alpha_cutoff}
\Gamma_\alpha \,\simeq\,\Gamma_{\alpha}^{(0)}
\left(\frac{\Lambda_0}{\Lambda_c^\alpha}\right)^{g_{\alpha}}\;.
\end{equation}
To write this equation in terms of invariant energy scales, we introduce
the strong coupling scale
\begin{equation}
\label{eq:T_K}
T_K \,\equiv\, \Gamma|_{V=\epsilon_0=0}\;,
\end{equation}
and write $\Gamma_\alpha$ in the form
\begin{equation}
\label{eq:gamma_alpha_invariant}
\Gamma_\alpha = T_K^\alpha 
\left({T_K\over\Lambda_c^\alpha}\right)^{g_\alpha}\;,
\end{equation}
with the independent scales
\begin{equation}
\label{eq:T_K_alpha}
T_K^\alpha \,\equiv\, \Gamma_\alpha^{(0)}
\left(\Lambda_0\over T_K\right)^{g_\alpha}\;.
\end{equation}
The scaling limit is defined by $\Gamma_\alpha^{(0)}\rightarrow 0$ and
$\Lambda_0\rightarrow\infty$, such that $T_K^\alpha$ remains constant. From
\eqref{eq:gamma_alpha_cutoff} and \eqref{eq:T_K} we see that $T_K$ is
determined from the self-consistent equation
\begin{equation}
\label{eq:T_K_sc}
T_K = \sum_\alpha T_K^\alpha = 
\sum_\alpha \Gamma_\alpha^{(0)}\left({\Lambda_0\over T_K}\right)^{g_\alpha}
\end{equation}
and remains also constant in the scaling limit. For symmetric Coulomb
interactions $g_\alpha=g$, we obtain the solution
\begin{equation}
\label{eq:T_K_symmetric}
T_K = \Gamma^{(0)}
\left({\Lambda_0\over\Gamma^{(0)}}\right)^{g\over 1+g}\;,\quad
T_K^\alpha = {\Gamma_\alpha^{(0)}\over\Gamma^{(0)}}T_K\;,
\end{equation}
with $\Gamma^{(0)}=\sum_\alpha\Gamma^{(0)}_\alpha$.

In the special case of two reservoirs $\alpha=L/R$, we use in the following
instead of $T_K^\alpha$ the invariant $T_K=T_K^L+T_K^R$ and the asymmetry
parameter $c^2=T_K^L/T_K^R$. We obtain
\begin{equation}
\label{eq:c_scale}
T_K^L={c^2\over 1+c^2}\,T_K\;,\quad T_K^R={1\over 1+c^2}\,T_K\;,
\end{equation}
and for symmetric Coulomb interactions
\begin{equation}
\label{eq:c_symmetric}
c\,=\,\sqrt{\Gamma^{(0)}_L\over\Gamma^{(0)}_R}\;.
\end{equation}

$T_K$ is the energy scale which determines the importance of charge
fluctuations.  Away from resonances, where $|\mu_\alpha-\epsilon_0|\gg
T_K$, charge fluctuations are weak and the RG flow of $\Gamma_\alpha$
is cut off by the scale $|\mu_\alpha-\epsilon_0|$, which describes the
distance to the resonance. Close to resonances, where
$|\mu_\alpha-\epsilon_0|\sim T_K$, charge fluctuations are strong, and
$\Gamma_\alpha$ is cut off by $T_K$.  Nevertheless, $\Gamma_\alpha$ is
bounded by the scale $T_K^\alpha$ for arbitrary system parameters even
for $V=\epsilon_0=0$, leading to finite results for all
cases. Although there is no rigorous argument why our theory should be
well-controlled in the presence of a single energy scale $T_K$, we
show in the next sections that in the scaling limit our results for
the charge susceptibility and the current are in excellent agreement
with exact numerical methods, provided that $U_\alpha^{(0)}\ll 1$.
This indicates that strong charge fluctuations are covered by our
theory. 

Close to resonance, where $\mu_\alpha\simeq\epsilon_0$, the rate
$\Gamma_\alpha$ is logarithmically enhanced, similar to corresponding
logarithmic enhancements for $2$-level models with spin fluctuations (Kondo
model), see Ref.~\onlinecite{RTRGKondo}. Defining an overall cutoff scale by
$\Lambda_c = \max\{\Lambda_c^L,\Lambda_c^R\}$ and expanding in $g_\alpha$, we
find close to the resonance
\begin{equation}
\label{eq:log_irlm}
\Gamma_\alpha\simeq\Gamma_\alpha^{(0)}
\left(\frac{\Lambda_0}{\Lambda_{\mathrm{c}}}\right)^{g_\alpha}
\left(1+g_\alpha\ln{\frac{\Lambda_{\mathrm{c}}}
{|\mu_\alpha-\epsilon_0+i\Gamma/2|}}\right)\;.
\end{equation}
In comparison to the Kondo model the IRLM is simpler in the sense that the
leading order charge fluctuation processes provide a unique cutoff scale
$|\mu_\alpha-\epsilon_0+i\Gamma/2|$ for the rates. In contrast, for the Kondo
model, the distance to the resonance as well as the Zeeman splitting itself
serve as cutoff parameters, such that different logarithms can occur for the
rates, see Ref.~\onlinecite{RTRGKondo} for details.

The appearance of a negative differential conductance in the IRLM at
large bias voltages (see Figs.~\ref{fig:fig6} and~\ref{fig:fig5}) can
be understood\cite{ndc} very easily from the form of the rates
$\Gamma_\alpha$, while from the numerical or field-theoretical
computation of the $I-V$ characteristics it is difficult to extract
the physical mechanisms.  In the limit $V\gg \Gamma,|\epsilon_0|$ the
current \eqref{eq:curr} for two reservoirs reduces to $I^{st}(V)\simeq
\frac{\Gamma_L\Gamma_R}{\Gamma}$, and to $I^{st}\simeq
(\Gamma_L^{(0)}\Gamma_R^{(0)}/(\Gamma^{(0)})^2)\Gamma$ for symmetric
Coulomb interactions $g_L=g_R=g$. Substituting the above expression
(\ref{eq:gamma_alpha_cutoff}) for $\Gamma$ being cut off by the
voltage, and using \eqref{eq:T_K_symmetric}, we obtain
\begin{equation}
I^{st}(V) =  {\Gamma_L^{(0)}\Gamma_R^{(0)}\over{\Gamma^{(0)}}^2} 
T_K\left({T_K\over V}\right)^{g}\sim V^{-g}\;,
\label{eq:IVpowerlaw}
\end{equation}
leading to a negative differential conductance for repulsive interactions.
The power-law behavior \eqref{eq:IVpowerlaw} was previously obtained using a
variety of other methods~\cite{Doyon,BSS,karrasch10,SA,ndc}. In contrast,
Nishino et al.\cite{NIH} find a critical value of $U=2$ above which negative
differential conductance appears. For attractive interactions, we obtain a
power-law increase of the current as a function of voltage which is consistent
with DMRG results in Ref.~\onlinecite{BSS}.  However, we will show in
Section~\ref{sec:current} that this result does no longer hold for asymmetric
Coulomb interactions.

\subsubsection{Charge susceptibility}

\begin{figure}[t!]
  \centering
\includegraphics[width=8.5cm,clip=true]{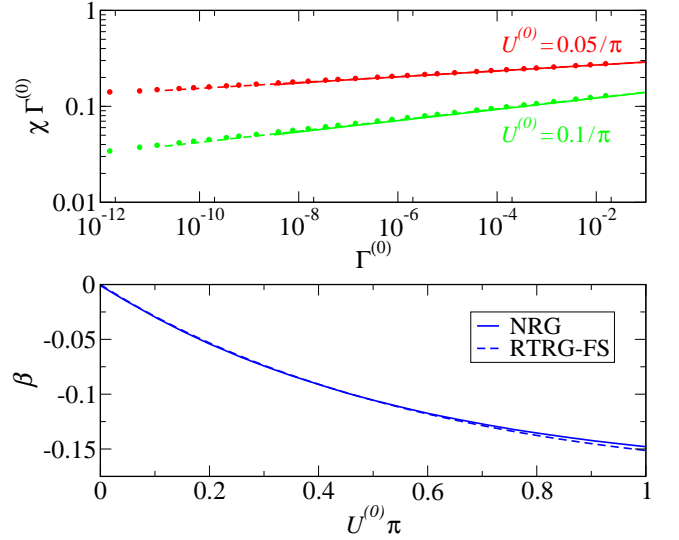}
\caption{\label{fig:fig4} (Color online)
  Results for the static susceptibility $\chi$ for the symmetric model with
  $U^{(0)}_L=U^{(0)}_R=U^{(0)}$ and
  $\Gamma^{(0)}_L=\Gamma^{(0)}_R=\Gamma^{(0)}$ at $\epsilon_0=V=0$;
  \emph{upper panel:} comparison of RTRG-FS (solid lines) results, the
  analytic solution of the flow equations (dashed lines), and NRG data
  (symbols); \emph{lower panel:} exponent $\beta=-g/(1+g)$.}
\end{figure}

The stationary charge susceptibility $\chi$ (or the static capacitance)
describes the charge response of the dot due to a shift of the level position
$\epsilon_0$ and is defined by
\begin{equation}
\label{eq:chi_def}
\chi\,=\,-{\partial n^{st} \over \partial\epsilon_0}\;.
\end{equation}
It can be obtained directly from \eqref{eq:n_general} and for arbitrary
level position reads
\begin{equation}
\label{eq:chi_general}
\chi\,=\,{1\over 2\pi}\sum_\alpha{\Gamma_\alpha\over (\mu_\alpha-\epsilon_0)^2
+({\Gamma\over 2})^2}\;,
\end{equation}
where we have neglected small corrections from the weak dependence of
$\Gamma_\alpha$ on $\epsilon_0$ via the power law
\eqref{eq:gamma_alpha_invariant}.  For the special case of two reservoirs with
$\mu_L=-\mu_R=V/2$ and for $\epsilon_0=0$, this gives
\begin{equation}
\label{eq:chi_special}
\chi|_{\epsilon_0=0}\,=\,{2\over\pi}{\Gamma\over V^2+\Gamma^2}\;.
\end{equation}
In particular at $V=0$, this result can be compared to exact numerical results
from NRG, which are shown in Fig.~\ref{fig:fig4}. We obtain
\begin{equation}
\label{eq:chi_equi}
\chi|_{\epsilon_0=V=0}\,=\,{2\over \pi\Gamma|_{\epsilon_0=V=0}}\,=
\,{2\over \pi T_K}\;,
\end{equation}
which can be used to define the physical scale $T_K$ even away from the
scaling limit.  For symmetric Coulomb interactions $g_L=g_R=g$, we can insert
$T_K$ from \eqref{eq:T_K_symmetric} and get
\begin{equation}
\label{eq:chi_equi_symmetric}
\chi|_{\epsilon_0=V=0}\,=\,{2\over \pi\Gamma^{(0)}}
\left({\Lambda_0\over\Gamma^{(0)}}\right)^
{-{g\over 1+g}}\;.
\end{equation}
As can be seen from Fig.~\ref{fig:fig4}, the exponent agrees
surprisingly well with the exact numerical result from NRG.  Since
$\epsilon_0=V=0$ is the most critical regime where strong charge
fluctuations are present, this comparison strongly supports that
our general solution \eqref{eq:chi_general} for arbitrary voltage and
arbitrary level position is a very good analytical approximation to
the exact result.

For $|{V\over 2}\pm \epsilon_0| \gg \Gamma$, power laws occur as function of
$V$ or $\epsilon_0$. From \eqref{eq:chi_general},
\eqref{eq:gamma_alpha_invariant} and \eqref{eq:c_scale} we obtain
\begin{eqnarray}
\label{eq:chi_off_resonance}
\chi &=& {1\over 2\pi}{c\over 1+c^2}\,T_K\\
\nonumber
&&\hspace{-1cm}\;\,\;\times\!\left(
{c\over ({V\over 2}-\epsilon_0)^2}
\left({T_K\over |{V\over 2}-\epsilon_0|}\right)^{g_L}\! \!\!+
{1\over c}{1\over ({V\over 2}+\epsilon_0)^2}
\left({T_K\over |{V\over 2}+\epsilon_0|}\right)^{g_R}\right).
\end{eqnarray}
For the symmetric case $g_L=g_R=g$ this leads to
\begin{equation}
\label{eq:chi_V}
\chi\,=\,{2\over\pi}{1\over T_K}\left({T_K\over V}\right)^{2+g}
\end{equation}
for $V\gg |\epsilon_0|$, and to
\begin{equation}
\label{eq:chi_epsilon}
\chi\,=\,{1\over 2\pi}{1\over T_K}\left({T_K\over |\epsilon_0|}\right)^{2+g}
\end{equation}
for $V\ll |\epsilon_0|$.

We note that for asymmetric Coulomb interactions the general result
\eqref{eq:chi_general} does not exhibit a clear power law if the
system is coupled to more than one reservoir, neither as function of
$V$ nor of $\epsilon_0$ nor of $\Gamma^{(0)}$. In this case, a linear
combination of different power laws is involved which does not reveal
a clear exponent except for if one of the energy scales is much larger
than the other two. 

\subsubsection{Current}\label{sec:current}

\begin{figure}[t!]
  \centering
\includegraphics[width=8.5cm,clip=true]{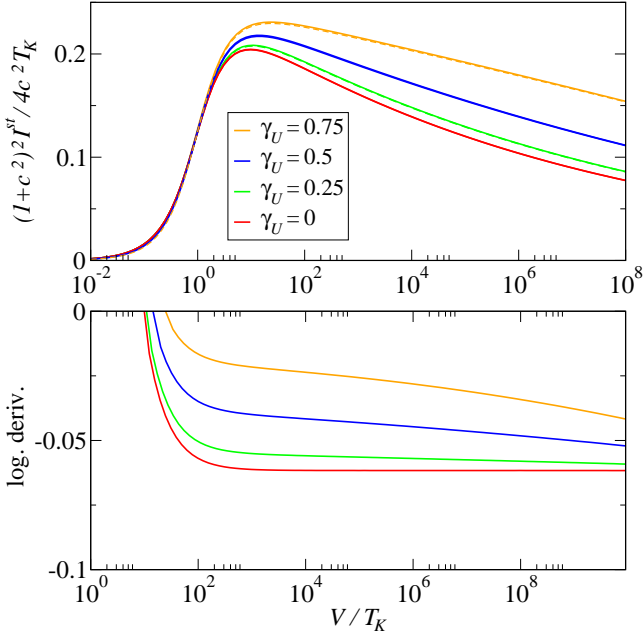}
\caption{\label{fig:fig6} (Color online)
  Results for the current $I(V)$ for asymmetric Coulomb interactions 
  $U_{L/R}^{(0)}=(1\pm \gamma_U) \, 0.1/\pi$ with $\gamma_U=0.75$, $0.5$, $0.25$, $0$ 
  from top to bottom and $t_L^{(0)}=t_R^{(0)}=0.001$, 
  $\epsilon_0=0$; the numerical solution (solid lines) is compared to the
  analytical result (dashed lines); \emph{lower panel:} logarithmic derivative.}
\end{figure}

In the case of two reservoirs the stationary current $I^{st}$ follows
from \eqref{eq:curr}.  We take $\epsilon_0>0$ and study the off- and
on-resonance case separately.  A comparison of the full numerical
solution of the flow equations to the analytical results obtained from
(\ref{eq:gamma_alpha_invariant}) and (\ref{eq:curr}) is provided in
Fig.~\ref{fig:fig6}. The excellent agreement shows that for this
situation already the simplified analytical treatment within poor
man's scaling yields an accurate description.

The {\it off-resonance} case is defined by $|V/2\pm\epsilon_0|\gg\Gamma$.
Using \eqref{eq:curr} we obtain
\begin{eqnarray}
\label{eq:current_off_large_V}
I^{st}&=&{\Gamma_L\Gamma_R\over\Gamma}\qquad\qquad\quad\,\,\,\,
\text{for}\quad{V\over 2}>\epsilon_0\quad,\\
\label{eq:current_off_small_V}
I^{st}&=&G_0 {\Gamma_L\Gamma_R\over \epsilon_0^2 - (V/2)^2}\,V
\quad\text{for}\quad{V\over 2}<\epsilon_0\quad.
\end{eqnarray}
Inserting for the rates from \eqref{eq:gamma_alpha_invariant} and
\eqref{eq:c_scale}, this gives
\begin{equation}
\label{eq:current_off_large_V_explicit}
I^{st}={c \over 1+c^2}T_K
{\left({T_K\over|V/2-\epsilon_0|}\right)^{g_L}
\left({T_K\over|V/2+\epsilon_0|}\right)^{g_R}\over
c\left({T_K\over|V/2-\epsilon_0|}\right)^{g_L}+{1\over c}
\left({T_K\over|V/2+\epsilon_0|}\right)^{g_R}} 
\end{equation}
for $V/2>\epsilon_0$, and
\begin{eqnarray}
\label{eq:current_off_small_V_explicit}
I^{st}&=&G_0{c^2 \over (1+c^2)^2}{T_K^2\over \epsilon_0^2-(V/2)^2}\nonumber\\
&&\times \left({T_K\over|V/2-\epsilon_0|}\right)^{g_L}
\left({T_K\over|V/2+\epsilon_0|}\right)^{g_R} V
\end{eqnarray}
for $V/2<\epsilon_0$.

From these results one can see in what cases a power law can be
expected.  First, for large voltages $V\gg \epsilon_0$, a power law
can only be seen for the symmetric model $g_L=g_R=g$, in which case
$I^{st}\sim V^{-g}$, see Fig.~\ref{fig:fig6}. This is the same result
obtained also in earlier studies~\cite{Doyon,BSS} of the IRLM.
However, in all other cases where $g_L\neq g_R$ a power law cannot be
seen on realistic scales since the two terms in the denominator of
Eq.~(\ref{eq:current_off_large_V_explicit}) are typically of the same
order of magnitude. 
The scale at which a definite power law is recovered is given by the condition
$(T_K/V)^{|g_L-g_R|/2}\ll 1$ being extremely small for $|g_L-g_R|\ll 1$. 
In Fig.~\ref{fig:fig6} the above condition is not met and the asymptotic behavior 
is not observed.
Only if in addition to $g_L\neq g_R$ the asymmetry
in the bare rates is large ($c \ll 1$ or $c \gg 1$), the power-law
behavior of $I^{st}(V)$ is recovered (with exponents $g_L$ or $g_R$,
respectively).

Interestingly, for $V\ll\epsilon_0$, a power law also occurs in the
asymmetric case, since $\Gamma$ does not appear in the denominator of 
\eqref{eq:current_off_small_V_explicit}. In this case we obtain 
\begin{equation}
\label{eq:current_epsilon}
I^{st}=G_0{c^2 \over (1+c^2)^2}
\left({T_K\over|\epsilon_0|}\right)^{2+g_L+g_R}\,V\;,
\end{equation}
i.e. a power law with exponent $-(2+g_L+g_R)$ always appears as function of
the level position $\epsilon_0$ at fixed voltage.

\begin{figure}[t!]
  \centering
\includegraphics[width=8.5cm,clip=true]{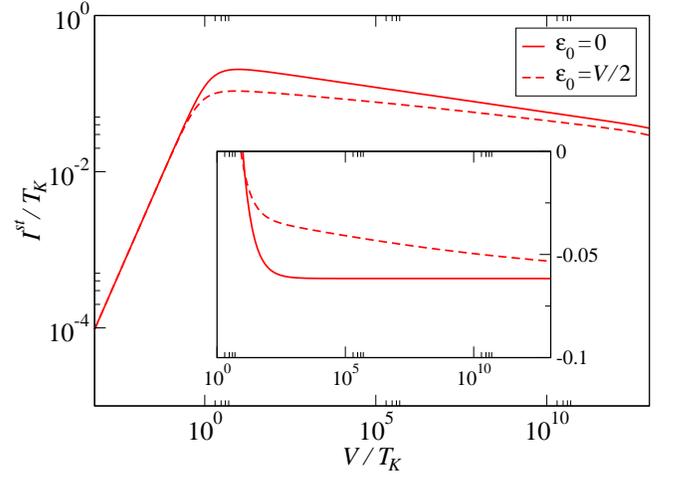}
\caption{\label{fig:fig5} (Color online)
 Results for the current $I(V)$ for the symmetric model with 
 $t_L^{(0)}=t_R^{(0)}=0.001$ and $U_L^{(0)}=U_R^{(0)}=0.1/\pi$ off-resonance 
 (solid lines) and on-resonance (dashed lines);
 \emph{inset:} logarithmic derivative.}
\end{figure}

In the {\it on-resonance} case $\epsilon_0=V/2$ the current is given by 
\begin{equation}
\label{eq:current_on}
I^{st}= \frac{\Gamma_L \Gamma_R}{2 \Gamma} 
= \frac{c}{1+c^2}\frac{T_K}{2}\frac{\left(\frac{T_K}{\Gamma}\right)^{g_L}
    \left(\frac{T_K}{V}\right)^{g_R}}
  {c\left(\frac{T_K}{\Gamma}\right)^{g_L}
    +\frac{1}{c}\left(\frac{T_K}{V}\right)^{g_R}}\;.
\end{equation}
It is important to note that if the level is in resonance with one of
the reservoirs it is not in resonance with the other one. Therefore,
at resonance the cutoff scales are $\Gamma$ for one rate and
$V$ for the other.  In contrast to the off-resonance case, no power
law appears even for the left-right symmetric model, see
Fig.~\ref{fig:fig5}. 
A power law is recovered only for unrealistically
large $V$, where the second term in the last denominator of
\eqref{eq:current_on} can be neglected leading to $I^{st}\sim V^{-g_R}
$. 
For the symmetric model shown in Fig.~\ref{fig:fig5} the condition 
$(\Gamma/V)^g = 0.01 \ll 1$ is fulfilled only for $V\sim 10^{30} T_K$.

A microscopic determination of the cutoff scales is therefore
essential to determine the correct on-resonance scaling behavior as a
function of the voltage, which does not simply appear as an additional
low-energy cutoff. The non-equilibrium physics for the generic
situation $\epsilon_0=\pm V/2$ and $g_L\neq g_R$ turns out to be more
complex and can not be inferred from the linear-response behavior.

\subsubsection{Conductance}

\begin{figure}[t!]
  \centering
\includegraphics[width=8.5cm,clip=true]{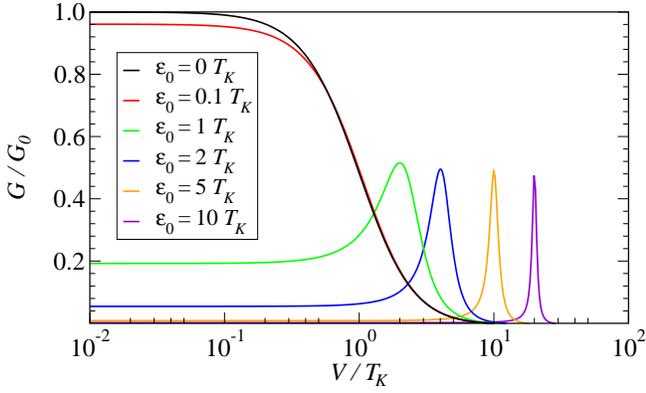}
\caption{\label{fig:fig7} (Color online)
 Conductance $G(V)=dI/dV$ for the symmetric model with
 $t_L^{(0)}=t_R^{(0)}=0.001$, $U^{(0)}_L=U^{(0)}_R=0.1/\pi$ and different
values of the gate voltage $\epsilon_0=0$, $0.1$, $1$, $2$, $5$, $10 \,T_K$ from top to bottom.}
\end{figure}

Another transport property of experimental interest is the conductance
$G={dI^{st}\over dV}$, which most vividly features the mentioned resonance
at $\epsilon_0=\pm V/2$ as the voltage becomes large, see Fig.~\ref{fig:fig7}.
Analytically, the conductance follows from differentiating \eqref{eq:curr}.
Neglecting small terms from the $V$-dependence of the rates $\Gamma_\alpha$,
we obtain
\begin{equation}
\label{eq:conductance}
G=G_0{2\Gamma_L\Gamma_R\over\Gamma^2}\left(
{({\Gamma\over 2})^2\over({V\over 2}-\epsilon_0)^2+({\Gamma\over 2})^2} +
{({\Gamma\over 2})^2\over({V\over 2}+\epsilon_0)^2+({\Gamma\over
2})^2}\right)\,,
\end{equation}
i.e. two Lorentzian resonances at $V/2=\pm\epsilon_0$.

\begin{figure}[t!]
  \centering
\includegraphics[width=8.5cm,clip=true]{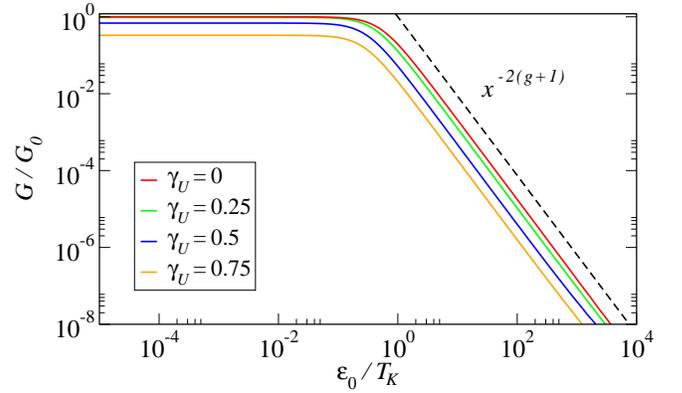}
\caption{\label{fig:fig7a} (Color online)
  Results for the conductance $G(\epsilon_0)$ for asymmetric Coulomb
  interactions $U_{L/R}^{(0)}=(1\pm \gamma_U) \, 0.1/\pi$ 
  with $\gamma_U=0$, $0.25$, $0.5$, $0.75$ from top to bottom and
  $t_L^{(0)}=t_R^{(0)}=0.001$ at $V=0$; the obtained power law is characterized by
  the exponent $\sim 2(g+1)$ in linear order, with $g=2U^{(0)}$.}
\end{figure}

For the off-resonance case $V,\Gamma\ll|\epsilon_0|$ we obtain
\begin{equation}
\label{eq:conductance_off_small_V}
G=G_0{\Gamma_L\Gamma_R\over \epsilon_0^2}=G_0
{c^2\over (1+c^2)^2}\left({T_K\over |\epsilon_0|}\right)^{2+g_L+g_R}
\end{equation}
in agreement with \eqref{eq:current_epsilon}.  Results from the
solution of the full flow equations are shown in Fig.~\ref{fig:fig7a}.

On the other hand, as a function of $V$ the current is given by 
\begin{equation}
\label{eq:curr_expansion}
I^{st}=G_0{2\Gamma_L\Gamma_R\over\Gamma}\left(\pi-2{\Gamma\over V}\right)
\end{equation}
for $V\gg\Gamma,|\epsilon_0|$, where we took into account the first correction
to the expansion of the arctan-function in \eqref{eq:curr}.
Interestingly, the latter leads to an additional regime characterized by a power law 
independently of the asymmetry.
Whereas the first term of Eq.~(\ref{eq:curr_expansion}) does not show a power law for
asymmetric Coulomb interactions (since $\Gamma=\Gamma_L+\Gamma_R$ appears in
the denominator), the second term does show a power law because $\Gamma$
cancels out. Taking the derivative with respect to $V$, the 
conductance reads
\begin{equation}
\label{eq:cond_expansion}
G=G_0\left({\partial\over\partial V}{2\Gamma_L\Gamma_R\over\Gamma}\right)
\left(\pi-2{\Gamma\over
V}\right)+G_0{2\Gamma_L\Gamma_R\over\Gamma}\left({2\Gamma\over
V^2}\right)\;.
\end{equation}
The derivative in the first term yields a factor $\sim g/V$ from the weak
voltage dependence of the rates $\Gamma_{L/R}$, which has to be compared with
the factor $\Gamma/V^2$ in the second term. Thus, for $g\ll \Gamma/V$
the second term dominates and yields a power law as function of the voltage
\begin{equation}
\label{eq:conductance_off_large_V}
G=G_0{4\Gamma_L\Gamma_R\over V^2}=G_0
{4c^2\over (1+c^2)^2}\left({T_K\over V}\right)^{2+g_L+g_R}\;.
\end{equation}
Thus, in contrast to the current, the conductance
shows always a power law either for large voltage or for large level position
with the same exponent $-(2+g_L+g_R)$ in the 
range $\Gamma,|\epsilon_0|\ll V\ll \Gamma/g$.

However, for $V\gg\Gamma/g$ the first contribution in \eqref{eq:curr_expansion} 
dominates leading to a negative differential conductance.

\subsection{Time evolution}\label{sec:time}

The time evolution of the reduced density matrix can be obtained directly from 
\eqref{eq:rhoDLeff} via inverse Laplace transform
\begin{equation}
  \rho_D(t)=\frac{i}{2\pi}\int\nolimits_{-\infty+i0^+}^{\infty+i0^+}dz\,
  \frac{e^{-izt}}{z-L_D^{eff}(z)}\,\rho_D(0)\;,
  \label{eq:timegeneralrho}
\end{equation}
where we have set the initial time to $t_0=0$.  We recall that we
assume the initial density matrix of the full system to be of the
product form \eqref{eq:rhoinitial}. This situation can be prepared by
setting the couplings between the leads and the dot to zero for times
$t<0$. At $t=0$ the couplings are suddenly switched on and the system
evolves under the Hamiltonian \eqref{eq:model}, which results in
\eqref{eq:timegeneralrho} for the time evolution of the reduced
density matrix. In conventional Markov approximation
$L_D^{eff}(z)\approx L_D^{eff}(z=0)$ one neglects the $z$-dependence
of the Liouvillian, which yields simple exponential decay towards the
stationary reduced density matrix. In contrast, we keep the
$z$-dependence of the Liouvillian including its branch cuts. The time
evolution of the current is obtained similarly by inverse Laplace
transform of \eqref{eq:Current}. This approach has previously been
used to study the real-time dynamics of the magnetization and current
in the anisotropic Kondo model~\cite{PSS}.

Specifically, using \eqref{eq:p_01} we find for the occupation
$n(t)=p_1(t)$ of the dot
\begin{equation}
  n(t)=\langle c^\dagger c\rangle(t)=\bigl(1+J_+(t)+J_-(t)\bigr)n(0)-J_+(t)\;,
\label{eq:dot_occup_int}
\end{equation}
where the auxiliary functions $J_\pm(t)$ are defined as
\begin{equation}
  J_\pm(t)=\frac{1}{2\pi}\int\nolimits_{-\infty+i0^+}^{\infty+i0^+}\frac{dz}{z}
  \frac{e^{-izt}}{z+i\Gamma(z)}\,\Gamma_\pm(z)\;.
  \label{eq:timeint1}
\end{equation}
Here $\Gamma_\pm(z)=\frac{1}{2}\Gamma(z)\pm\Gamma'(z)$ with
$\Gamma(z)=\sum_\alpha\Gamma_\alpha(z)$ and $\Gamma'(z)=\sum_\alpha
\Gamma_\alpha'(z)$ are obtained from the analytic continuation of
\eqref{eq:solution_gamma}, \eqref{eq:solution_gamma_prime} and 
\eqref{eq:gamma_tilde_final} (see below).  Similarly, using 
\eqref{eq:curr_laplace} we find for the current in lead $\gamma$ 
\begin{eqnarray}
  I_\gamma(t)&=&\label{eq:timeint2}\\
  &&\!\!\!\!\!\!\!\!\!\!\frac{i}{2\pi}\!\!\!
  \int\limits_{-\infty+i 0^+}^{\infty+i 0^+}\!\!\!
  \frac{dz}{z}\frac{e^{-i zt}}{\Gamma(z)}
  \bigl(\Gamma^1_\gamma(z)\Gamma_-(z)+\Gamma^2_\gamma(z)\Gamma_+(z)\bigr)
  \nonumber\\*
  & &\!\!\!\!\!\!\!\!\!\!+\frac{i}{2\pi}\!\!\!
  \int\limits_{-\infty+i 0^+}^{\infty+i 0^+}\!\!\!dz
  \frac{e^{-i zt}}{\Gamma(z)}
  \frac{\Gamma^1_\gamma(z)\!-\!\Gamma^2_\gamma(z)}{z+i\Gamma(z)}
  \bigl(\Gamma_+(z)\!-\!n(0)\Gamma(z)\bigr)\;,\nonumber
\end{eqnarray}
where
$\Gamma_\gamma^{1/2}(z)=\Gamma_\gamma'(z)\pm\frac{1}{2}\Gamma_\gamma(z)$.

In order to evaluate the dot occupation and the current in the interacting
case we start with the analytic continuation of
Eqs.~\eqref{eq:solution_gamma}, \eqref{eq:solution_gamma_prime},
\eqref{eq:gamma_epsilon_final} and \eqref{eq:gamma_tilde_final}. We consider
the case of two reservoirs with $\mu_L=-\mu_R=V/2$ and restrict
ourselves to the situation of symmetric couplings to the leads, i.e.
$U_L^{(0)}=U_R^{(0)}=U^{(0)}$ and $\Gamma_L^{(0)}=\Gamma_R^{(0)}\equiv
\Gamma^{(0)}/2$. In this case we can make use of the helpful identity
$\Gamma_\epsilon(-E)^*=\Gamma_\epsilon(E)$, which follows from
\eqref{eq:gamma_property} and \eqref{eq:gamma_epsilon_final}.  As a result the
analytic continuation reads
\begin{eqnarray}
  \Gamma_\epsilon(z)&=&\frac{\Gamma^{(0)}}{2}\sum_\alpha 
  \left(\frac{\Lambda_0}{\Gamma(z-\mu_\alpha)
      -i(z-\mu_\alpha)}\right)^{g}\;,\label{eq:timeGamma1}\\
  \Gamma_\alpha(z)&=&\frac{\Gamma^{(0)}}{4}
  \left[\left(\frac{\Lambda_0}{\frac{1}{2}\Gamma_\epsilon(z+\mu_\alpha)
      -i(z+\mu_\alpha-\epsilon_0)}\right)^{g}\right.\nonumber\\
   &&\;\!\!\!\left.+\left(\frac{\Lambda_0}
       {\frac{1}{2}\Gamma_\epsilon(z-\mu_\alpha)
      -i(z-\mu_\alpha+\epsilon_0)}\right)^{g}
  \right]\;,\label{eq:timeGamma2}\\
  \Gamma'_\alpha(z)&=&-\frac{i\Gamma^{(0)}}{8\pi U^{(0)}}
  \left[\left(\frac{\Lambda_0}{\frac{1}{2}\Gamma_\epsilon(z+\mu_\alpha)
      -i(z+\mu_\alpha-\epsilon_0)}\right)^{g}\right.\nonumber\\
   &&\;\!\!\!\left.-\left(\frac{\Lambda_0}
       {\frac{1}{2}\Gamma_\epsilon(z-\mu_\alpha)
      -i(z-\mu_\alpha+\epsilon_0)}\right)^{g}
  \right]\;,
  \label{eq:timeGamma3}
\end{eqnarray}
with $g=2U^{(0)}(1-U^{(0)})$. To proceed we first calculate the
non-vanishing poles $z_1\equiv-i\tilde{\Gamma}$ and
$z_\pm\equiv\pm\tilde{\epsilon}-\frac{i}{2}\tilde{\Gamma}_\epsilon$ of
the resolvent $1/(z-L_D^{eff}(z))$. Using the parameterization
\eqref{eq:L_form} we find $-i\Gamma(z_1)=z_1$, $\epsilon(z_+)=z_+$,
and $-\epsilon(-z_-^*)=z_-^*$, which results in
\begin{equation}
  \Gamma(-i\tilde{\Gamma})=\tilde{\Gamma}\;,\quad
  \Gamma_\epsilon(\pm\tilde{\epsilon}-\tfrac{i}{2}\tilde{\Gamma}_\epsilon)=
  \tilde{\Gamma}_\epsilon\pm 2i(\tilde{\epsilon}-\epsilon_0)\;.
  \label{eq:timepoles}
\end{equation}
Second, we approximate the functions $\Gamma(z)$ and
$\Gamma_\epsilon(z)$ in the denominators of
\eqref{eq:timeGamma1}--\eqref{eq:timeGamma3} by their fixed points,
i.e. we replace
\begin{equation}
  \Gamma(z)\rightarrow\tilde{\Gamma}\;,\quad
  \Gamma_\epsilon(z)\rightarrow\tilde{\Gamma}_\epsilon
  \pm 2i(\tilde{\epsilon}-\epsilon_0)\;.
  \label{eq:timeapp}
\end{equation}
We use the upper (lower) approximation for $\Gamma_\epsilon(z)$ in the terms
with a singularity at $z\approx\epsilon_0$ ($z\approx-\epsilon_0$). Inserting
\eqref{eq:timeapp} we obtain
\begin{eqnarray}
  \Gamma_\epsilon(z)&=&\frac{T_K}{2}\sum_\alpha 
  \left(\frac{T_K}{\tilde{\Gamma}
      -i(z-\mu_\alpha)}\right)^{g}\;,\label{eq:timeGammase}\\
  \Gamma_\alpha(z)&=&\frac{T_K}{4}
  \left[\left(\frac{T_K}{\frac{1}{2}\tilde{\Gamma}_\epsilon
      -i(z+\mu_\alpha-\tilde{\epsilon})}\right)^{g}\right.\nonumber\\
   &&\left.
     +\left(\frac{T_K}{\frac{1}{2}\tilde{\Gamma}_\epsilon
      -i(z-\mu_\alpha+\tilde{\epsilon})}\right)^{g}
  \right]\;,\\
  \Gamma'_\alpha(z)&=&-\frac{iT_K}{8\pi U^{(0)}}
  \left[\left(\frac{T_K}{\frac{1}{2}\tilde{\Gamma}_\epsilon
      -i(z+\mu_\alpha-\tilde{\epsilon})}\right)^{g}\right.\nonumber\\
   &&\left.
     -\left(\frac{T_K}{\frac{1}{2}\tilde{\Gamma}_\epsilon
      -i(z-\mu_\alpha+\tilde{\epsilon})}\right)^{g}
  \right]\nonumber\\
     &\simeq&-\frac{i}{2\pi}\Gamma_\alpha(z)\,
      \ln\frac{\frac{1}{2}\tilde{\Gamma}_\epsilon
      -i(z-\mu_\alpha+\tilde{\epsilon})}{\frac{1}{2}\tilde{\Gamma}_\epsilon
      -i(z+\mu_\alpha-\tilde{\epsilon})},
  \label{eq:timeGammasSL}
\end{eqnarray}
where have expanded the prefactor in $\Gamma_\alpha'(z)$ up to $O(1)$ while 
keeping the exponents unchanged. We note that the consistent calculation of 
$O(U^{(0)})$ corrections of the prefactor in $\Gamma_\alpha'(z)$ 
requires the analysis of three-loop diagrams 
which is beyond the scope of our work. Furthermore 
we have taken the scaling limit $\Gamma^{(0)}\rightarrow
0$, $\Lambda_0\rightarrow\infty$ with
$T_K^{1+g}=\Gamma^{(0)}\Lambda_0^g$ kept constant.  The decay rates
$\tilde{\Gamma}$ and $\tilde{\Gamma}_\epsilon$ and the renormalized
level position $\tilde{\epsilon}$ can be evaluated numerically from
Eq.~\eqref{eq:timepoles}.  The physical interpretation of these
quantities is as follows: Whereas $\tilde{\Gamma}$ describes the
charge relaxation processes on the dot and thus the relaxation of the
diagonal elements of the reduced density matrix with respect to the
charge states, $\tilde{\Gamma}_\epsilon$ is the broadening of the
local level induced by the coupling to the leads, i.e.  it
characterizes the relaxation of the off-diagonal
elements. Furthermore, the coupling to the leads yields a
renormalization of the level position from the bare value $\epsilon_0$
to $\tilde{\epsilon}$. For weak Coulomb interactions, $U^{(0)}\le
0.1$, this renormalization is found to be small,
$|\tilde{\epsilon}/\epsilon_0|\le 0.01$.  The approximate analytical
expressions \eqref{eq:timeGammase}--\eqref{eq:timeGammasSL} show
excellent agreement with the full numerical solution of the RG
equations (\ref{eq:rg_gamma_tilde})-(\ref{eq:rg_chi}).

\begin{figure}[t]
  \centering
  \includegraphics[width=80mm,clip=true]{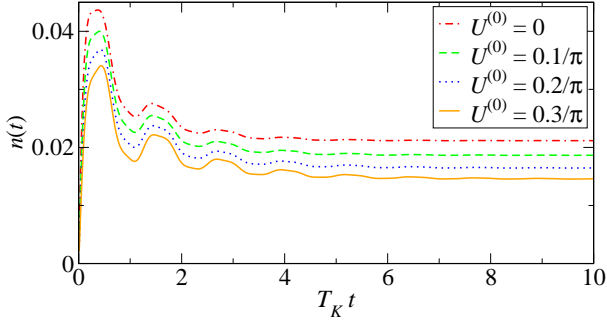}
  \caption{(Color online) Time evolution of the dot occupation $n(t)$
    for $V=\epsilon_0=10\,T_K$ and different values of $U^{(0)}$. The initial
    condition is given by $n(0)=0$. We observe oscillating behavior at short
    times, $T_Kt\le 5$.} 
  \label{fig:fig10}
\end{figure}

\begin{figure}[t]
  \centering
  \includegraphics[width=80mm,clip=true]{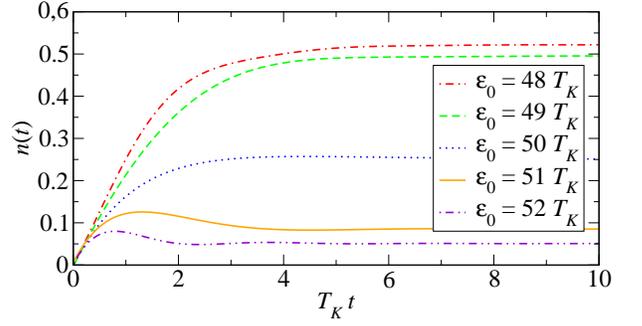}
  \caption{(Color online) Time evolution of the dot occupation $n(t)$
    for $U^{(0)}=0.1/\pi$, $V=100\,T_K$ and different values of $\epsilon_0$. 
    The initial condition is given by $n(0)=0$.}
  \label{fig:fig11}
\end{figure}

\begin{figure}[t]
  \centering
  \includegraphics[width=80mm,clip=true]{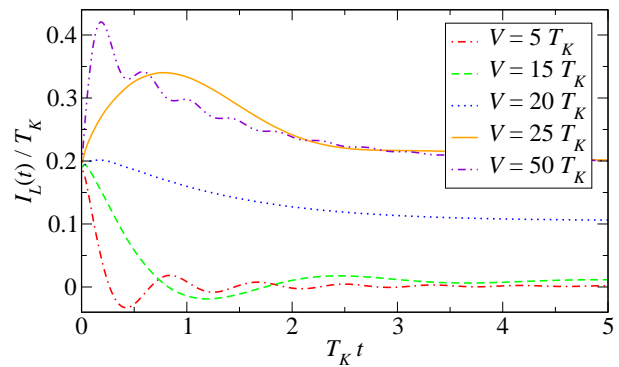}
  \caption{(Color online) Time evolution of the current $I_L(t)$ in
    the left lead for $U^{(0)}=0.1/\pi$, $\epsilon_0=10\,T_K$ and different
    values of $V$.  The initial condition is given by $n(0)=0$. The
    non-zero current at $t=0$ is determined by the displacement
    current $dn(t)/dt$. The current oscillates with frequency $\sim
    \epsilon_0-V/2$; note the absence of oscillations on resonance.}
  \label{fig:fig12}
\end{figure}

Inspecting the integral representations \eqref{eq:timeint1} and
\eqref{eq:timeint2} we see that the dominant contributions stem from the
singularities in the lower half-plane of the involved functions.  We stress
that the approximations \eqref{eq:timeapp} preserve
this analytic structure, i.e. the poles as well as the positions and exponents
of the branch cuts from the power laws remain unchanged. The integrals
\eqref{eq:timeint1} and \eqref{eq:timeint2} can then be treated using standard
techniques of contour integrations (see App.~\ref{app:time}).  Numerical
evaluation yields the occupation of the dot $n(t)$ as well as the current
$I_L(t)$ in the left lead shown in Figs.~\ref{fig:fig10}--\ref{fig:fig12}.  

Furthermore, for the long-time behavior off resonance
($|\epsilon_0-V/2|\gg T_K, 1/t$) we are able to derive approximate
analytical expressions for the dot occupation~\cite{endnote3}
\begin{eqnarray}
   n(t)&\approx& n^{st}-\left(\frac{1}{2}+
      \frac{\Gamma'(-i\tilde{\Gamma})}{\tilde{\Gamma}}\right)
    e^{-\tilde{\Gamma}t}\nonumber\\
    &&+\,
    \frac{(T_Kt)^{1+g}}{2\pi}\,e^{-\tilde{\Gamma}_\epsilon t/2}\,
  \left[\frac{\sin\left((\tilde{\epsilon}\!+\!\tfrac{V}{2})t\right)}
    {(\tilde{\epsilon}\!+\!\tfrac{V}{2})^2\,t^2}\right.\nonumber\\
  &&\left.-\,\pi U\frac{\cos\left((\tilde{\epsilon}\!+
         \!\tfrac{V}{2})t\right)}
      {(\tilde{\epsilon}\!+\!\tfrac{V}{2})^2\,t^2} + (V\!\to\!-V)\right]\!
\label{eq:ntresult}
\end{eqnarray}
as well as for the current (we assume $V\gg T_K, 1/t$ in addition)
\begin{eqnarray}
   I_L(t)&\approx& I_L^{st}+\Gamma_L(-i\tilde{\Gamma})\left(
    \frac{1}{2}+\frac{\Gamma'(-i\tilde{\Gamma})}{\tilde{\Gamma}}\right)
    e^{-\tilde{\Gamma}t}\nonumber\\
    &&+
    \frac{T_K}{2\pi}(T_Kt)^{g}\,e^{-\tilde{\Gamma}_\epsilon t/2}\,
  \frac{\cos\left((\tilde{\epsilon}\!-\!\tfrac{V}{2})t\right)}
      {(\tilde{\epsilon}\!-\!\tfrac{V}{2})\,t}\;,\label{eq:Itresult}
\end{eqnarray} 
with the stationary values given by \eqref{eq:p_stationary} [or \eqref{eq:n}] and
\eqref{eq:I_stationary}, respectively. For simplicity we have
considered the level to be initially empty, $n(0)=0$. We note that in \eqref{eq:ntresult}
the prefactor of the sine is determined up to $O(1)$ while the prefactor of the 
cosine has been calculated consistently up to $O(U^{(0)})$. The cosine represents the 
leading oscillatory behavior for $\epsilon_0=0$.  From
Figs.~\ref{fig:fig10}--\ref{fig:fig12} as well as \eqref{eq:ntresult}
and \eqref{eq:Itresult} we observe that the time evolution is governed
by an exponential decay towards the stationary values, characterized
by the decay rates $\tilde{\Gamma}$ and $\tilde{\Gamma}_\epsilon/2$.
In addition, oscillating terms with explicitly voltage-dependent
frequencies $\tilde{\epsilon}\pm V/2$ appear, accompanied by an
interaction-dependent power-law decay $\sim t^{g-1}$. The last result
is of particular importance for applications in error correction
schemes of quantum information processing as it violates the standard
assumption~\cite{ECS} of a purely exponential decay.  The same
qualitative features were observed for the time evolution in the
anisotropic Kondo model~\cite{PSS}. We stress that these qualitative
features are independent of the approximations leading to
\eqref{eq:timeGammase}--\eqref{eq:timeGammasSL} as they are completely
determined by the analytic structure. The imaginary parts of poles and
branch points lead to exponential decay, their real parts yield
oscillating behavior, and the integrations along the branch cuts
result in power laws.

In Fig.~\ref{fig:fig12} we observe that the current starts at a non-zero
value. This is due to a non-vanishing displacement
current~\cite{displacementI} $dn(t)/dt$, i.e. the fluctuating number
of particles on the dot. Specifically, the particle number
conservation in the full system implies
\begin{equation}
I_L(t)+I_R(t)=\frac{dn(t)}{dt}\;,
\end{equation} 
where $I_{L/R}(t)=-dN_{L/R}/dt$ is the current flowing out of lead
$L/R$. The initial condition $n(0)$ chosen in Fig.~\ref{fig:fig12} causes
particles to flow from the leads to the dot as soon as the couplings
$t^{(0)}_{L/R}$ are switched on. These initial currents establish on
the time scale $t\sim 1/D$ with $D$ denoting the band width, i.e. they
start instantaneously in the scaling limit. The strong charge
fluctuations on the dot further result in situations where particles
flow off the dot into the leads even against the applied bias voltage,
as can be seen by the appearance of $I_L(t)<0$ in Fig.~\ref{fig:fig12}.
A similar displacement current has been observed by Schmidt et
al.~\cite{Schmidtea} in the transient dynamics of the Anderson
impurity model, where also the effects of different reservoir cut-offs
have been investigated.

In the non-interacting case $U^{(0)}=0$ the rates and the level position are
simply given by $\tilde{\Gamma}=\tilde{\Gamma}_\epsilon=T_K=\Gamma^{(0)}$, and
$\tilde{\epsilon}=\epsilon_0$. The contour integrals \eqref{eq:timeint1} and
\eqref{eq:timeint2} can be evaluated explicitly (see App.~\ref{app:time}),
resulting in 
\begin{equation}
 n (t) = e^{-T_Kt} n (0) + \frac{1 - e^{-T_Kt}}{2} + F_0 (t) - F_1 (t) 
\label{eq:n_NI} 
\end{equation}
and
\begin{eqnarray}
I_{\gamma} (t) &=& T_K e^{-T_Kt} \frac{1 - 2 n (0)}{4}  \nonumber \\
&+&\frac{T_K}{2} \left[ F_{0,\gamma} (t) - F_{0,\bar{\gamma}} (t) \right] 
+\frac{T_K}{2} F_1 (t) ,
\label{eq:I_NI}
\end{eqnarray}
where $\bar{\gamma} = - \gamma$ and $F_{0/1} (t) = \sum_{\alpha} F_{0/1,
  \alpha} (t)$. The final expression for $F_{0/1, \alpha} (t)$ is given by the
formula
\begin{eqnarray}
 F_{0/1, \alpha} (t) &=& \frac{e^{-T_Kt/2 \pm T_Kt/2}}{2 \pi} \,
\left[ \pm \arctan \frac{\mu_{\alpha} -\epsilon_0}{T_K/2} \right. 
\nonumber \\
& & \left. \hspace{-1.85cm}+\, \mathrm{Im} \, \mathrm{E}_1 \!
\left(\pm \frac{T_Kt}{2} - i (\epsilon_0 - \mu_{\alpha}) t \right) \!+\!\frac{\pi}{2}(1\mp 1)\mathrm{sign} (\mu_{\alpha}-\epsilon_0) \right] \!\nonumber\\
&&
\label{eq:F01main}
\end{eqnarray}
In the stationary limit $t \to \infty$ we recover from \eqref{eq:n_NI} and
\eqref{eq:I_NI} the stationary values \eqref{eq:n} and \eqref{eq:curr},
respectively. In the opposite limit, at $t=0^+$, we use the property
\eqref{eq:F01_0} and observe a non-zero initial value of the current
\begin{equation} 
I_{\gamma} (t=0^+) = T_K\frac{1 - 2 n (0)}{4}\;,
\end{equation}
i.e. the displacement current discussed above. We note that
\eqref{eq:n_NI}--\eqref{eq:F01main} contain exponentially decaying terms with
rates $T_K$ and $T_K/2$, oscillations with frequencies $\epsilon_0\pm V/2$,
and power-law behavior $\sim 1/t$. 

Finally, we would like to compare our results for the non-interacting
model \eqref{eq:n_NI}--\eqref{eq:F01main} with the literature. As is
well-known~\cite{Leggett} there exists a mapping (in a certain
parameter regime) between the resonant level model (and thus the
anisotropic Kondo model) and the spin-boson model (or double-well
problem) of dissipative quantum mechanics. By studying the time
evolution in the latter, Lesage and Saleur~\cite{LesageSaleur} showed
that generically one has to expect relaxation with various decay rates
as well as oscillating terms in qualitative agreement with our
results.  Later Anders and Schiller~\cite{AndersSchiller} addressed
the time evolution in the resonant level model. They derived analytic
results for the dot occupation $n(t)$ in the single-lead model, which
are identical to the $V\to0$ limit of \eqref{eq:n_NI}. They further
considered quite general initial density matrices beyond the product
form \eqref{eq:rhoinitial} and observed a decay with two different
relaxation rates accompanied by algebraic decay as well as
oscillations with frequency $\epsilon_0$ in this more general setting
as well.  Komnik~\cite{Komnik} extended their results to the two-lead
model and studied the current through the system. The results he
obtained are similar to \eqref{eq:n_NI} and
\eqref{eq:I_NI}. Furthermore, time evolution and quench dynamics in
the resonant level model have been studied in the context of the
anisotropic Kondo model at the Toulouse point~\cite{HeylKehrein}.

\section{Conclusion}

We presented a non-equilibrium RG scheme to study transport properties of
quantum dots in the regime of strong charge fluctuations. We developed a
gauge-invariant approximation scheme to solve the RG equations analytically by
expanding all quantities around zero Matsubara frequency. We illustrated the
approach by a minimal and nontrivial model: the IRLM. Whereas many previous
works treated the problem numerically or at the self-dual point, our
analytical treatment in the scaling limit and for moderate Coulomb
interactions reveals the renormalized tunneling rates
$\Gamma_\alpha$ parametrizing the occupation and the current in the same form as the
noninteracting solution~\cite{}. The tunneling rates are given by a power law cut off by the
distance to resonances. At resonance the total tunneling rate
$\Gamma=\sum_\alpha\Gamma_\alpha$ itself is the cutoff scale leading to a
self-consistent equation for $\Gamma$. We calculated the power-law exponents
up to second order in the Coulomb interaction and found a very accurate
agreement with NRG. Each $\Gamma_\alpha$ has its own power-law exponent
$g_\alpha=2U_\alpha^{(0)}-\sum_\beta (U_\beta^{(0)})^2$ determined by the
Coulomb interaction $U_\alpha^{(0)}$ between the dot and reservoir $\alpha$.
As already pointed out in Ref.~\onlinecite{SA} it turned out that the current
does not reveal power laws as function of the voltage in the generic case of
asymmetric Coulomb interactions. The reason is that an asymmetry factor
$\Gamma_L\Gamma_R/\Gamma$ occurs, which contains a linear combination of all
rates in the denominator. In contrast, away from the particle-hole symmetric
point, we found that power laws occur as function of the level position in the
generic case, since, for large level position, only a factor
$\Gamma_L\Gamma_R$ appears for the current or the linear conductance.
However, the charge susceptibility shows a power law neither as function
of the voltage nor of the level position in the presence of more than one
reservoir, since it contains a sum of terms, each being proportional to the
rate $\Gamma_\alpha$.

Whereas the RTRG-FS scheme is limited to the scaling limit, the functional RG
allows to access the steady state for arbitrary tunneling parameters beyond
the scaling limit. As shown in Ref.~\onlinecite{SA}, both methods provide
excellent agreement in the scaling limit. The combined use of both RG
approaches provides hence a complete picture of the non-equilibrium physics
under consideration. As shown in this paper, an advantage of the RTRG-FS
method is the analytic treatment of Coulomb interactions up to next-to-leading
order, providing an excellent agreement of power-law exponents with NRG
results. Furthermore, the time evolution can be studied with RTRG-FS, where we
found complex relaxation dynamics similarly to previous studies of the
dynamics of the non-equilibrium Kondo model~\cite{PSS}.

The understanding of basic models of spin and charge fluctuations opens the
way for applications to more complex quantum dot models.  A fundamental issue
for the future concerns the universality of the effects of strong charge
fluctuations at resonances found for the IRLM where they induce a level
broadening and a renormalization of the tunneling couplings.  In particular,
in the presence of both spin and charge fluctuations, as e.g. in the
non-equilibrium Anderson model, the level position itself becomes renormalized
and it is still an open question what the precise line shape of resonances
looks like. Whereas strong charge fluctuations at resonances seem to be
described by the RTRG-FS method, an open question remains whether strong spin
or orbital fluctuations can be covered as well. In both cases, a single energy
scale dominates the physics and cuts off the RG flow. Surprisingly, although
no rigorous argument allows the truncation of the RG equations in this case,
for strong charge fluctuations within the IRLM we have shown here that our
results agree very accurately with exact numerical methods. Whether such an
agreement holds also for strong spin and orbital fluctuations will be studied
in future works.

\section{Acknowledgments}

We thank N. Andrei, B. Doyon, C. Karrasch, D. Kennes, V. Meden, P. Schmitteckert, A.
Tsvelik, and A. Zawadowski for discussions.  This work was supported by the
DFG-FG 723 and 912, the Robert Bosch Foundation, and by the AHV.

\appendix
\section{Derivation of the RG equations}
\label{app:flow} 

In this Section we report a detailed derivation of the flow equations
and their evaluation.  As outlined in detail in Ref.~\onlinecite{S},
the RG consists of two steps: The first one is a discrete step where
the symmetric part of the reservoir Fermi function is integrated out,
and the second step a continuous RG transformation where the Matsubara
frequencies of the reservoir Fermi function are integrated out
successively (note that at $T=0$ the Matsubara frequencies are
continuous).  In addition to Ref.~\onlinecite{S} we will present here
a systematic treatment of the frequency dependence of the vertices,
which can be quite generically used for the treatment of strong charge
fluctuations. In contrast to Rfs.~\onlinecite{S,RTRGKondo}, where a
systematic weak coupling expansion around the poor man scaling
solution has been presented, we propose here a systematic expansion
around the point where all Matsubara frequencies are set to zero.  The
dependence on the Laplace variable $E$ is fully taken into account,
leading to a gauge-invariant theory. Many considerations presented
here hold in general and can also be used for other models in the
charge fluctuation regime. In particular, we will make use of various
generic cancellations of diagrams which simplify the RG analysis
considerably.

\begin{figure}[t]
\center{\includegraphics{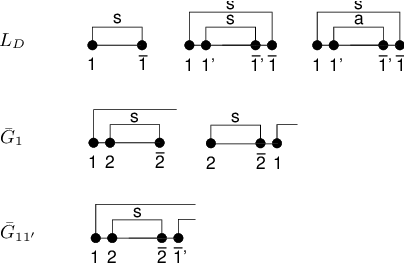}}
\caption{\label{fig:fig2}
 Diagrams of the discrete RG step. Double vertices are represented by 
 dots lying close to each other. The indices $s/a$ refer to the 
 symmetric/antisymmetric part of the reservoir contraction.}
\end{figure}

We consider a model with single and double vertices, described by the
interaction \eqref{eq:interaction} initially. In Liouville space the
vertices are defined in
(\ref{eq:single_vertices})-(\ref{eq:double_vertices_bt}). Following
Ref.~\onlinecite{S}, we start with the discrete RG step and integrate
out the symmetric part of the reservoir Fermi function by a
perturbative treatment.  The respective diagrams are shown in
Fig.~\ref{fig:fig2} and define the initial values of the second
continuous RG procedure.  For the initial Liouvillian we obtain
\begin{eqnarray}
  L_D(E)&=&L_D^{(0)} -  i\frac{\pi}{2}\bar{G}^{(0)}_1\tilde{G}^{(0)}_{\bar{1}}
  -i\frac{\pi^2}{16}D\bar{G}^{(0)}_{11'}\bar{G}^{(0)}_{\bar{1}'\bar{1}}
  \nonumber\\
  &&+\frac{\pi^2}{32}\bar{G}^{(0)}_{11'}(E_{11'}-L_D^{(0)})\bar{G}^{(0)}_{\bar{1}'\bar{1}}
  -\frac{\pi}{4}D\bar{G}^{(0)}_{11'}\tilde{G}^{(0)}_{\bar{1}'\bar{1}}
  \nonumber \\ 
  &  &- i\frac{\pi}{4}\bar{G}^{(0)}_{11'}(E_{11'}-
  L_D^{(0)})\tilde{G}^{(0)}_{\bar{1}'\bar{1}}\;,\phantom{\frac{\pi^2}{32}}
\end{eqnarray}
where we used the short-hand notation $E_{1\dots
  n}=E+\bar{\mu}_1+\dots+\bar{\mu}_n$ with
$\bar{\mu}_i=\eta_i\mu_{\alpha_i}$.  The reservoir band width $D$ is
related in a certain way to the initial value $\Lambda_0$ of the
continuous RG flow, see Eq.~\eqref{eq:D_initial_cutoff} below.

The initial vertices are given by 
\begin{eqnarray}
  \bar{G}_1=\bar{G}_1^{(0)} - i\frac{\pi}{2}\bar{G}^{(0)}_{12}\tilde{G}^{(0)}_{\bar{2}}
  - i\frac{\pi}{2}\bar{G}^{(0)}_{2}\tilde{G}^{(0)}_{\bar{2}1}
\end{eqnarray}
and
\begin{eqnarray}
  \bar{G}_{11'}=\bar{G}_{11'}^{(0)} - i\frac{\pi}{2}(\bar{G}^{(0)}_{12}\tilde{G}^{(0)}_{\bar{2}1'}
  - \bar{G}^{(0)}_{1'2}\tilde{G}^{(0)}_{\bar{2}1})\;.
\end{eqnarray}
The equations for $\Sigma_{\gamma}$ and $\bar{I}_1^{\gamma}$ are
determined analogously, the first vertex just has to be replaced by
$\bar{I}^{\gamma(0)}_1$. We note that $\bar{I}_{12}^{\gamma(0)}=0$ for
our model where double vertices describe a Coulomb interaction between
the dot and the reservoirs, i.e. these processes can not contribute to
the current vertex. For other models, where double vertices describe
spin or orbital fluctuations, $\bar{I}_{12}^{\gamma(0)}$ has to be
included as well.

Inserting the form \eqref{eq:bare} of the initial matrices into above
expressions and comparing with the parametrizations
\eqref{eq:L_form}-\eqref{eq:ikernel} yield the following initial
values for the continuous RG flow
\begin{eqnarray}
\nonumber 
\Gamma_+ &=& \Gamma_-=\frac{1}{2}\Gamma_{\alpha}^{(0)}
\;,\quad \Gamma_{\alpha}^{(0)}=2\pi(t_{\alpha}^{(0)})^2\;,\\
\nonumber
\epsilon(E) &=&\epsilon_0(1-\frac{\pi^2}{16}(U_{\alpha}^{(0)})^2)+
E\frac{\pi^2}{16}(U_{\alpha}^{(0)})^2\\
\nonumber
&&-\frac{i}{2}\Gamma_{\alpha}^{(0)}-i\frac{\pi^2}{8}D(U_{\alpha}^{(0)})^2\;,\\
\nonumber
t_\alpha &=& t_{\alpha}^{(0)}\;,\quad 
t_{\alpha}^{\gamma} = \delta_{\alpha\gamma}t_{\alpha}^{(0)}\;,\\
\nonumber
t_\alpha^2 &=& t_{\alpha}^{(0)}-i\pi t_{\alpha}^{(0)}U_{\alpha}^{(0)}\;,\quad
t_\alpha^3=t_{\alpha}^{(0)}+i\pi t_{\alpha}^{(0)}U_{\alpha}^{(0)}\;,\\
\nonumber
\Gamma_{\gamma}^1 &=& \frac{1}{2}\Gamma_{\gamma}^{(0)} \;,\quad
\Gamma_{\gamma}^2=-\frac{1}{2}\Gamma_{\gamma}^{(0)}\;,\\
\label{eq:initial_values}
U_{\alpha} &=& U_{\alpha}^{(0)}\;.
\end{eqnarray}

We proceed with the flow equations for the continuous RG procedure.
The Laplace variable is decomposed into real and imaginary part as
$z=E+i\omega\equiv (E,\omega)$. The Liouvillian $L_D(E,\omega)$ and
the vertices $\bar{G}_1(E,\omega;\omega_1)$ and
$\bar{G}_{11'}(E,\omega;\omega_1,\omega_1^\prime)$ acquire an
additional dependence on the Laplace variable and on Matsubara
frequencies $\omega_1$ and $\omega_1^\prime$. The diagrams taken into
account are shown in Fig.~\ref{fig:fig3}. We consider contributions to
the flow of $L_D$, $\bar{G}_1$ and $\bar{G}_{11'}$ to lowest order in
$\Gamma$ and to next-to-leading order in $U_{\alpha}$ to describe the
scaling limit and to obtain exponents up to order
$O(U_{\alpha}^2)$. Terms of order $\sim \Gamma U_{\alpha}$ for
$\bar{G}_{11'}$ are neglected. These would generate nonzero elements
in the upper left $2\times2$ block of \eqref{eq:vert2}.

\begin{widetext}
\begin{figure}[ht]
  \centering
\includegraphics[scale=1]{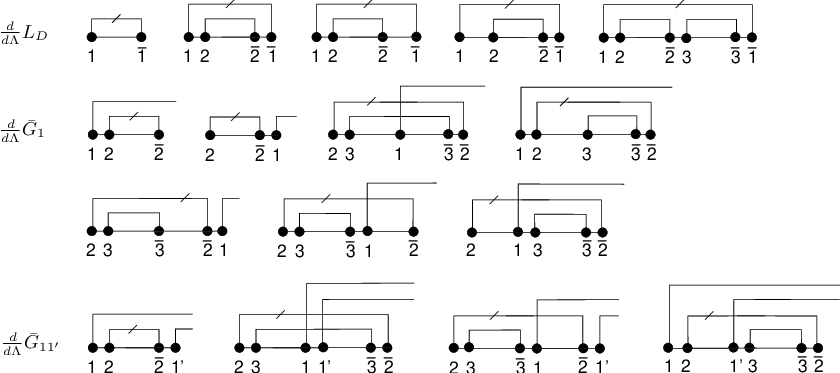}
\caption{\label{fig:fig3} 
  Diagrams of the RG equations. Double vertices are represented by dots lying
  close to each other. For $\bar{G}_{11'}$ diagrams have to be subtracted
  where the indices $1$ and $1'$ are interchanged, provided this gives a new
  diagram. This guarantees the relation $\bar{G}_{11'}=-\bar{G}_{1'1}$.}
\end{figure}
Using the diagrammatic rules developed in Ref.~\onlinecite{S}, the RG
equations for the Liouvillian and the vertices read
\begin{eqnarray}
\nonumber
- \frac{d}{d \Lambda} L_D (E,\omega) &=& 
-i \, \bar{G}_1(E,\omega;\Lambda) \, \Pi(E_1,\omega+\Lambda) \, \bar{G}_{\bar{1}}(E_1,\omega+\Lambda;-\Lambda)\\
\nonumber && 
+\,(-i)^2 \, \bar{G}_{12}(E,\omega;\Lambda,\omega_2) \, \Pi(E_{12},\omega+\Lambda+\omega_2) \,
\bar{G}_{\bar{2}\bar{1}}(E_{12},\omega+\Lambda+\omega_2;-\omega_2,-\Lambda) \\
\nonumber &&
+\,(-i)^2 \, \bar{G}_{12}(E) \, \Pi(E_{12},\omega+\Lambda+\omega_2) \, \bar{G}_{\bar{2}}(E_{12}) \, 
\Pi(E_1,\omega+\Lambda) \, \bar{G}_{\bar{1}}(E_1)\\
\nonumber &&
+\,(-i)^2 \, \bar{G}_1(E) \, \Pi(E_1,\omega+\Lambda) \, \bar{G}_2(E_1) \, 
\Pi(E_{12},\omega+\Lambda+\omega_2) \, \bar{G}_{\bar{2}\bar{1}}(E_{12})\\
\label{eq:L_rg} &&
+\,(-i)^3 \, \bar{G}_{12}(E) \, \Pi(E_{12},\omega+\Lambda+\omega_2) \, \bar{G}_{\bar{2}3} (E_{12}) \,
\Pi(E_{13},\omega+\Lambda+\omega_3) \, \bar{G}_{\bar{3}\bar{1}}(E_{13})\;,\\
\nonumber&&\\
\nonumber 
- \frac{d}{d \Lambda} \bar{G}_1(E,\omega;\omega_1) &=& 
-i \, \bar{G}_{12}(E,\omega;\omega_1,\Lambda) \, \Pi(E_{12},\omega+\omega_1+\Lambda) \,
\bar{G}_{\bar{2}}(E_{12},\omega+\omega_1+\Lambda;-\Lambda)\\
\nonumber &&
-i \, \bar{G}_2(E,\omega;\Lambda) \, \Pi(E_2,\omega+\Lambda) \, 
\bar{G}_{\bar{2}1}(E_2,\omega+\Lambda;-\Lambda,\omega_1)\\
\nonumber &&
+\,(-i)^2 \, \bar{G}_{23}(E) \, \Pi(E_{23},\omega+\Lambda+\omega_3) \, \bar{G}_{1}(E_{23}) \, 
\Pi(E_{123},\omega+\omega_1+\Lambda+\omega_3) \, \bar{G}_{\bar{3}\bar{2}}(E_{123})\\
\nonumber &&
+\,(-i)^2 \, \bar{G}_{12}(E) \, \Pi(E_{12},\omega+\omega_1+\Lambda) \, \bar{G}_{3}(E_{12}) \, 
\Pi(E_{123},\omega+\omega_1+\Lambda+\omega_3) \, \bar{G}_{\bar{3}\bar{2}}(E_{123})\\
\nonumber &&
+\,(-i)^2 \, \bar{G}_{23}(E) \, \Pi(E_{23},\omega+\Lambda+\omega_3) \, \bar{G}_{\bar{3}}(E_{23}) \, 
\Pi(E_{2},\omega+\Lambda) \, \bar{G}_{\bar{2}1}(E_{2})\\
\nonumber &&
-\,(-i)^2 \, \bar{G}_{23}(E) \, \Pi(E_{23},\omega+\Lambda+\omega_3) \, \bar{G}_{\bar{3}1}(E_{23}) \, 
\Pi(E_{12},\omega+\omega_1+\Lambda) \, \bar{G}_{\bar{2}}(E_{12})\\
\label{eq:G_1_rg} &&
-\,(-i)^2 \, \bar{G}_{2}(E) \, \Pi(E_{2},\omega+\Lambda) \, \bar{G}_{13}(E_{2}) \, 
\Pi(E_{123},\omega+\omega_1+\Lambda+\omega_3) \, \bar{G}_{\bar{3}\bar{2}}(E_{123})\;,\\
\nonumber &&\\
\nonumber
- \frac{d}{d \Lambda} \bar{G}_{11'} (E,\omega;\omega_1,\omega_1^\prime) &=&
-i \, \bar{G}_{12}(E,\omega;\omega_1,\Lambda) \, \Pi(E_{12},\omega+\omega_1+\Lambda) \, 
\bar{G}_{\bar{2}1'}(E_{12},\omega+\omega_1+\Lambda;-\Lambda,\omega_1^\prime)\\
\nonumber &&
+i \, \bar{G}_{1'2}(E,\omega;\omega_1^\prime,\Lambda) \, \Pi(E_{1'2},\omega+\omega_1^\prime+\Lambda) \, 
\bar{G}_{\bar{2}1}(E_{1'2},\omega+\omega_1^\prime+\Lambda;-\Lambda,\omega_1)\\
\nonumber &&
+\,(-i)^2 \, \bar{G}_{23}(E) \, \Pi(E_{23},\omega+\Lambda+\omega_3) \, 
\bar{G}_{11'}(E_{23}) \, \Pi(E_{11'23},\omega+\omega_1+\omega_1^\prime+\Lambda+\omega_3) \, 
\bar{G}_{\bar{3}\bar{2}}(E_{11'23})\\
\nonumber &&
-\,(-i)^2 \, \bar{G}_{23}(E) \, \Pi(E_{23},\omega+\Lambda+\omega_3) \, 
\bar{G}_{\bar{3}1}(E_{23}) \, \Pi(E_{12},\omega+\omega_1+\Lambda) \, 
\bar{G}_{\bar{2}1'}(E_{12})\\
\nonumber &&
+\,(-i)^2 \, \bar{G}_{23}(E) \, \Pi(E_{23},\omega+\Lambda+\omega_3) \, 
\bar{G}_{\bar{3}1'}(E_{23}) \, \Pi(E_{1'2},\omega+\omega_1^\prime+\Lambda) \, 
\bar{G}_{\bar{2}1}(E_{1'2})\\
\nonumber &&
-\,(-i)^2 \, \bar{G}_{12}(E) \, \Pi(E_{12},\omega+\omega_1+\Lambda) \, 
\bar{G}_{1'3}(E_{12}) \, \Pi(E_{11'23},\omega+\omega_1+\omega_1^\prime+\Lambda+\omega_3) \, 
\bar{G}_{\bar{3}\bar{2}}(E_{11'23})\\
\nonumber &&
+\,(-i)^2 \, \bar{G}_{1'2}(E) \, \Pi(E_{1'2},\omega+\omega_1^\prime+\Lambda) \, 
\bar{G}_{13}(E_{1'2}) \, \Pi(E_{11'23},\omega+\omega_1+\omega_1^\prime+\Lambda+\omega_3) \, 
\bar{G}_{\bar{3}\bar{2}}(E_{11'23})\;,\\
\label{eq:G_2_rg} 
\end{eqnarray}
\end{widetext}
where 
\begin{equation}
\label{eq:Pi}
\Pi(E,\omega)\,=\,{1\over E+i\omega-L_D(E,\omega)}
\end{equation}
and 
\begin{equation}
\label{eq:G_zero}
\bar{G}_1(E)\equiv\bar{G}_1(E,0;0)\;,\quad
\bar{G}_{11'}(E)\equiv\bar{G}_{11'}(E,0;0,0)\;.
\end{equation}
Implicitly, one has to sum over all indices on the r.h.s. of
\eqref{eq:L_rg}-\eqref{eq:G_2_rg}, which do not appear on the l.h.s. In
addition, one has to perform the integral $\int_0^\Lambda d\omega_2$ and
$\int_0^\Lambda d\omega_3$ at all places where the frequencies $\omega_{2/3}$
occur. The RG equations for the current kernel and the current vertex are
analogous to \eqref{eq:L_rg} and \eqref{eq:G_1_rg}, respectively, the only
difference is that the first vertex has to be replaced by the current vertex.

Except for the real part $E$ of the Laplace variable, all frequencies are
bounded by the cutoff $\Lambda$. Since $\Lambda\rightarrow 0$ finally, it is
natural to account for the dependence on the Matsubara frequencies by
expanding the vertices around the reference value \eqref{eq:G_zero}, where all
Matsubara frequencies are set to zero. Therefore, we have neglected the
frequency dependence of the vertices in the higher order terms of
\eqref{eq:L_rg}-\eqref{eq:G_2_rg}. The frequency dependence of the vertices is
calculated in leading order by neglecting the frequency dependence of the
vertices on the r.h.s.  of \eqref{eq:G_1_rg} and \eqref{eq:G_2_rg} together
with omitting the higher order terms in these equations. This gives
\begin{widetext}
\begin{eqnarray}
\nonumber 
\frac{d}{d \Lambda} \left\{\bar{G}_1(E,\omega;\omega_1)-\bar{G}_1(E)\right\} &\simeq& 
i \, \bar{G}_{12}(E) \, \left(\Pi(E_{12},\omega+\omega_1+\Lambda)-\Pi(E_{12},\Lambda)\right) \,
\bar{G}_{\bar{2}}(E_{12})\\
\label{eq:G_1_leading_rg} &&
+i \, \bar{G}_2(E) \, \left(\Pi(E_2,\omega+\Lambda)-\Pi(E_2,\Lambda)\right) \, 
\bar{G}_{\bar{2}1}(E_2)\;,\\
\nonumber &&\\
\nonumber
\frac{d}{d \Lambda} \left\{\bar{G}_{11'} (E,\omega;\omega_1,\omega_1^\prime)-\bar{G}_{11'}(E)\right\} &\simeq&
i \, \bar{G}_{12}(E) \, \left(\Pi(E_{12},\omega+\omega_1+\Lambda)-\Pi(E_{12},\Lambda)\right) \, 
\bar{G}_{\bar{2}1'}(E_{12})\\
\label{eq:G_2_leading_rg} &&
-i \, \bar{G}_{1'2}(E) \, \left(\Pi(E_{1'2},\omega+\omega_1^\prime+\Lambda)-\Pi(E_{1'2},\Lambda)\right) \, 
\bar{G}_{\bar{2}1}(E_{1'2})\;.
\end{eqnarray}
\end{widetext}
To integrate these equations in leading order we first define a function
$F(E,\omega)$ by
\begin{equation}
\label{eq:F}
i\Pi(E,\omega)\,=\,{d\over d\omega}F(E,\omega)\;.
\end{equation}
Neglecting the weak logarithmic $\Lambda$-dependence of $L_D(E,\omega)$
generated by the RG (note that
$\Pi(E,\omega)\equiv\Pi_\Lambda(E,\omega)$ depends implicitly on $\Lambda$ via
$L_D(E,\omega)$), we can use 
\begin{equation}
\label{eq:F_approx}
i\Pi(E,\omega+\Lambda)\,\simeq\,{d\over d\Lambda}F(E,\omega+\Lambda)\;.
\end{equation}
Using this in \eqref{eq:G_1_leading_rg} and \eqref{eq:G_2_leading_rg}, and
neglecting in addition the weak logarithmic $\Lambda$-dependence of the
vertices generated by RG, we can integrate these equations to
\begin{widetext}
\begin{eqnarray}
\nonumber 
\bar{G}_1(E,\omega;\omega_1)  &\simeq& \bar{G}_1(E) \,
+ \, \bar{G}_{12}(E) \, \left(F(E_{12},\omega+\omega_1+\Lambda)-F(E_{12},\Lambda)\right) \,
\bar{G}_{\bar{2}}(E_{12})\\
\label{eq:G_1_leading} &&
\qquad\qquad + \, \bar{G}_2(E) \, \left(F(E_2,\omega+\Lambda)-F(E_2,\Lambda)\right) \, 
\bar{G}_{\bar{2}1}(E_2)\;,\\
\nonumber &&\\
\nonumber
\bar{G}_{11'} (E,\omega;\omega_1,\omega_1^\prime) &\simeq& \bar{G}_{11'}(E)\,
+ \, \bar{G}_{12}(E) \, \left(F(E_{12},\omega+\omega_1+\Lambda)-F(E_{12},\Lambda)\right) \, 
\bar{G}_{\bar{2}1'}(E_{12})\\
\label{eq:G_2_leading} &&
\qquad\qquad - \, \bar{G}_{1'2}(E) \, \left(F(E_{1'2},\omega+\omega_1^\prime+\Lambda)-F(E_{1'2},\Lambda)\right) \, 
\bar{G}_{\bar{2}1}(E_{1'2})\;.
\end{eqnarray}
\end{widetext}

To find the RG equations for $\bar{G}_1(E)$, $\bar{G}_{11'}(E)$ and 
\begin{equation}
\label{eq:L_E}
L_D(E)\,=\,L_D(E,0)\;,
\end{equation}
we set $\omega=\omega_1=\omega_1^\prime=0$ in
\eqref{eq:L_rg}-\eqref{eq:G_2_rg}, and insert the results
\eqref{eq:G_1_leading} and \eqref{eq:G_2_leading} for the frequency dependence
of the vertices in the lowest order terms. Furthermore, the frequency
integrations can be performed by using
\begin{equation}
\label{eq:Pi_integration}
i\int_0^\Lambda d\omega \,\Pi(E,\omega+\Lambda)\,=\,K(E)\;,
\end{equation}
with 
\begin {equation}
\label{eq:K_function}
K(E)\,=\,F(E,2\Lambda)-F(E,\Lambda)\;.
\end{equation}
Collecting the various terms one finds after some straightforward algebra
\begin{widetext}
\begin{eqnarray}
\nonumber
\frac{d}{d\Lambda}L_D(E) \,&=&\, 
i \, \bar{G}_1(E) \, \Pi(E_1,\Lambda) \, \bar{G}_{\bar{1}}(E_1) \,
- \, i \, \bar{G}_{12}(E) \, K (E_{12}) \, \bar{G}_{\bar{2}\bar{1}}(E_{12})\\
\label{eq:L_final_rg} &&
- \, 2i \, \bar{G}_{12}(E) \, K (E_{12}) \, \bar{G}_{\bar{2}3}(E_{12}) \, K (E_{13}) \,
\bar{G}_{\bar{3}\bar{1}}(E_{13})\;,\\
\nonumber &&\\
\nonumber
\frac{d}{d\Lambda}\bar{G}_1(E) \,&=&\,
i \, \bar{G}_{12}(E) \, \Pi (E_{12},\Lambda) \, \bar{G}_{\bar{2}}(E_{12}) \,
+ \, i \, \bar{G}_{2}(E) \, \Pi (E_{2},\Lambda) \, \bar{G}_{\bar{2}1}(E_{2})\\
\label{eq:G_1_final_rg} &&
+ \, \bar{G}_{23}(E) \, \Pi (E_{23},\Lambda+\omega_3) \, \bar{G}_{1}(E_{23}) \,
\Pi (E_{123},\Lambda+\omega_3) \, \bar{G}_{\bar{3}\bar{2}}(E_{123})\;,\\
&&\nonumber\\
\nonumber
\frac{d}{d\Lambda}\bar{G}_{11'}(E) \,&=&\,
i \, \bar{G}_{12}(E) \, \Pi (E_{12},\Lambda) \, \bar{G}_{\bar{2}1'}(E_{12}) \,
- \, i \, \bar{G}_{1'2}(E) \, \Pi (E_{1'2},\Lambda) \, \bar{G}_{\bar{2}1}(E_{1'2})\\
\label{eq:G_2_final_rg} &&
+ \, \bar{G}_{23}(E) \, \Pi (E_{23},\Lambda+\omega_3) \, \bar{G}_{11'}(E_{23}) \,
\Pi (E_{11'23},\Lambda+\omega_3) \, \bar{G}_{\bar{3}\bar{2}}(E_{11'23})\;.
\end{eqnarray}
\end{widetext}
It turns out that many generic cancellations occur. In particular the
corrections from the frequency dependence of the vertices from the lowest
order terms cancel with corresponding diagrams in higher orders. For
$\bar{G}_1(E)$ ($\bar{G}_{11'}(E)$) only the first three (two) diagrams of
Fig.~\ref{fig:fig3} remain with frequency independent vertices. For $L_D(E)$
the cancellation is not complete but the third and fourth diagrams cancel
against frequency dependent corrections of the first two diagrams, whereas the
last diagram obtains a factor $2$.  These cancellations simplify the RG analysis
considerably and appear to be a generic model-independent feature.

To calculate the remaining frequency integrations in \eqref{eq:G_1_final_rg}
and \eqref{eq:G_2_final_rg}, and to find explicit representations for
$\Pi(E,\omega)$, $F(E)$ and $K(E)$, we first introduce the spectral
decomposition of the Liouvillian
\begin{equation}
\label{eq:L_spectral}
L_D(E,\omega)\,=\,\sum_i \lambda_i(E,\omega)\,P_i(E,\omega)\;.
\end{equation}
Here, $\lambda_i$ denote the eigenvalues of the Liouvillian and $P_i$ the
projectors onto the eigenstates (note that the Liouvillian is non-hermitian,
so that the eigenvalues are complex valued, and the right and left eigenvectors
are not identical).  \eqref{eq:L_spectral} leads to a corresponding spectral
representation of the resolvent $\Pi(E,\omega)$, defined in \eqref{eq:Pi}. In
leading order, we again expand in $\omega$ up to first order for the
eigenvalues $\lambda_i(E,\omega)$, whereas we neglect the $\omega$-dependence
of the projectors $P_i(E,\omega)$. This leads to the approximation
\begin{equation}
\label{eq:Pi_approx}
\Pi(E,\omega)\,\simeq\,-i\sum_i\,{Z_i(E)\over \omega-i\chi_i(E)}\,P_i(E)\;,
\end{equation}
where we defined the Z-factor
\begin{equation}
\label{eq:Z_factor}
Z_i(E)\,=\,{1\over 1-{d\over dE}\lambda_i(E)}
\end{equation}
and the distance to the resonant positions
\begin{equation}
\label{eq:chi_function}
\chi_i(E)\,=\,Z_i(E)(E-\lambda_i(E))\;.
\end{equation}
Here, $\lambda_i(E)\equiv\lambda_i(E,0)$ and $P_i(E)\equiv P_i(E,0)$.
Within this approximation we obtain
\begin{eqnarray}
\label{eq:F_function_approx}
\!\!\!\!\!\!\!\!\!\!\!F(E,\omega) &\simeq& \sum_i Z_i(E)\ln(\omega-i\chi_i(E))P_i(E)\;,\\
\label{eq:K_function_approx}
\!\!\!\!\!\!\!\!\!\!\!K(E) &\simeq& \sum_i Z_i(E)\ln\left({2\Lambda-i\chi_i(E)\over \Lambda-i\chi_i(E)}\right)P_i(E)\;.
\end{eqnarray}
The set of RG equations
\eqref{eq:L_final_rg}-\eqref{eq:G_2_final_rg} is thus complete and can be solved
for a specific model. In particular, it turns out that our approximations are
gauge-invariant, i.e. if all single-particle levels of the dot, all chemical
potentials of the reservoirs and the Laplace variable $E$ are shifted by the
same amount, all physical observables remain the same. This is only the case
if the $E$-dependence of all quantities is fully taken into account, an
expansion around a fixed value of $E$, like e.g. $E=0$, would not lead to a
gauge-invariant theory.

We now turn to the evaluation of the RG equations for the IRLM. Using the form
\eqref{eq:L_form} for the Liouvillian, the eigenvalues are given by
\begin{eqnarray}
\nonumber
\lambda_0(E)&=&0\;,\quad \lambda_1(E)=-i\Gamma(E)\;,\\
\label{eq:eigenvalues_irlm}
\lambda_+(E)&=&\epsilon(E)\;,\quad \lambda_-(E)=-\epsilon(-E)^*\;, 
\end{eqnarray}
with the corresponding projectors
\begin{eqnarray}
P_0(E)&=& \frac{1}{\Gamma(E)}\left( \begin{array}{cccc}
 \Gamma_-(E)& \Gamma_-(E)&0&0 \\
  \Gamma_+(E)&\Gamma_+(E) &0&0 \\
 0&0&0&0\\
0&0  &0&0\end{array} \right)\;,
\label{eq:proj}\nonumber\\
P_1(E)&=& \frac{1}{\Gamma(E)}\left( \begin{array}{cccc}
 \Gamma_+(E)& -\Gamma_-(E)&0&0 \\
  -\Gamma_+(E)&\Gamma_-(E) &0&0 \\
 0&0&0&0\\
0&0  &0&0\end{array} \right)\;,\nonumber\\
P_+(E)&= &\left( \begin{array}{cccc}
0& 0&0&0 \\
  0&0 &0&0 \\
 0&0&1&0\\
0&0  &0&0\end{array} \right)\;,\quad
P_-(E)= \left( \begin{array}{cccc}
0& 0&0&0 \\
 0&0&0&0 \\
 0&0&0&0\\
0&0  &0&1\end{array} \right)\;.\nonumber\\&&
\end{eqnarray}
Using $\Gamma(E)=\Gamma(-E)^*$, we obtain for the $Z$-factors and the
$\chi$-functions
\begin{eqnarray}
\label{eq:Z_1_irlm}
\!\!\!\!\!\!\!\!\!\!\!\!\!\!\!\!\!\!Z_1(E) &=& Z_1(-E)^* = (1+i{d\over dE}\Gamma(E))^{-1}\;,\\
\label{eq:chi_1_irlm}
\!\!\!\!\!\!\!\!\!\!\!\!\!\!\!\!\!\!\chi_1(E) &=& -\chi_1(-E)^* = Z_1(E)(E+i\Gamma(E))\;,\\
\label{eq:Z_irlm}
\!\!\!\!\!\!\!\!\!\!\!\!\!\!\!\!\!\!Z(E) &\equiv& Z_+(E) = Z_-(-E)^* = (1-{d\over dE}\epsilon(E))^{-1}\;,\\
\!\!\!\!\!\!\!\!\!\!\!\!\!\!\!\!\!\!\chi(E) &\equiv& \chi_+(E) =\chi_-(-E)^* = Z(E)(E-\epsilon(E))\;.
\label{eq:chi_irlm}
\end{eqnarray}

Summing over $\eta$ and taking into account
$\bar{G}_{+\alpha,-\alpha}(E)=-\bar{G}_{-\alpha,+\alpha}(E)$, the flow
equations \eqref{eq:L_final_rg}-\eqref{eq:G_2_final_rg} can be simplified to
\begin{widetext}
\begin{eqnarray}
\label{eq:rgl}
\frac{d}{d\Lambda}L_D(E)&=&
i\bar{G}_{+\alpha}(E)\Pi (E+\mu_{\alpha},\Lambda)\bar{G}_{-\alpha}(E+\mu_{\alpha})
+i\bar{G}_{-\alpha}(E)\Pi (E-\mu_{\alpha},\Lambda)\bar{G}_{+\alpha}(E-\mu_{\alpha})\nonumber \\
&&-2i\bar{G}_{+\alpha,-\alpha}(E) K(E)\bar{G}_{+\alpha,-\alpha}(E)\;,\phantom{\frac{1}{2}}\\
\label{eq:rgvert1}
\frac{d}{d\Lambda}\bar{G}_{+\alpha}(E)&=&
i\bar{G}_{+\alpha,-\alpha}(E)\Pi (E,\Lambda)\bar{G}_{+\alpha}(E)
-i\bar{G}_{+\alpha}(E)\Pi (E+\mu_{\alpha},\Lambda)\bar{G}_{+\alpha,-\alpha}(E+\mu_{\alpha})\;, \\
\label{eq:rgvert2}
\frac{d}{d\Lambda}\bar{G}_{+\alpha,-\alpha}(E)&=&
2\bar{G}_{+\alpha',-\alpha'}(E)\Pi (E,\Lambda+\omega_3)
\bar{G}_{+\alpha,-\alpha}(E)\Pi (E,\Lambda+\omega_3)\bar{G}_{+\alpha',-\alpha'}(E)\;.
\end{eqnarray}
\end{widetext}
The RG equations for the current kernel and vertex read
\begin{eqnarray}
 \frac{d}{d \Lambda} \Sigma_{\gamma} (E)\! &=& \! 
i \bar{I}^{\gamma}_{+ \alpha} (E) \Pi (E +\mu_{\alpha} 
+ i \Lambda) \bar{G}_{- \alpha} (E + \mu_{\alpha}) \nonumber\\&& \!
+ i \bar{I}_{- \alpha}^{\gamma} (E) \Pi (E -\mu_{\alpha} 
+ i \Lambda) \bar{G}_{+ \alpha} (E - \mu_{\alpha})\;, \phantom{\frac{1}{1}}\nonumber\\
 \frac{d}{d \Lambda} \bar{I}_{+ \alpha}^{\gamma} (E)\! &=& \!  
- i \bar{I}_{+\alpha}^{\gamma} (E) \Pi (E +\mu_{\alpha} 
+ i \Lambda ) \bar{G}_{+\alpha, -\alpha} (E +\mu_{\alpha})\;, \nonumber\\
 \frac{d}{d \Lambda} \bar{I}_{- \alpha}^{\gamma} (E)\! &=& \!   
i \bar{I}_{-\alpha}^{\gamma} (E) \Pi (E -\mu_{\alpha} 
+ i \Lambda ) \bar{G}_{+\alpha, -\alpha} (E +\mu_{\alpha})\;.\nonumber\\&&
\end{eqnarray}

For the derivation of the explicit flow equations for the effective parameters
as introduced in the matrix representations of the Liouvillian and the
vertices we use the following helpful identities:
\begin{eqnarray}
&&\bar{G}_{\eta\alpha}P_{-\eta}=P_{\eta}\bar{G}_{\eta\alpha}=0\;,\nonumber\\
&&\bar{G}_{+\alpha,-\alpha'}P_1=P_1\bar{G}_{+\alpha,-\alpha}=0\;,\nonumber\\
&&P_+\bar{G}_{-\alpha}P_-=P_-\bar{G}_{+\alpha}P_+=0\;,\nonumber\\
&&\bar{G}_{12}P_i\bar{G}_3P_j\bar{G}_{45}=0 \;\;\; {\rm  for} \;\;\;i,j=1,\pm\;,\nonumber\\
&&\bar{G}_{+\alpha}(E)
P_1(E_1)\bar{G}_{-\alpha'}=-(t_\alpha^2-t_\alpha^3)(E)t_1^{\alpha'}(-E')^*P_-\;,\nonumber\\
&&\bar{G}_{-\alpha}(E)P_1(E_1)\bar{G}_{+\alpha'}=(t_\alpha^2-t_\alpha^3)(-E)^*t_1^{\alpha'}(-E')P_+\;,\nonumber\\
&&\bar{G}_{+\alpha,-\alpha'}(E)P_+=P_+\bar{G}_{+\alpha,-\alpha}(E)=U_{\alpha}(E)P_+\;,\nonumber\\
&&\bar{G}_{+\alpha,-\alpha'}(E)P_-=P_+\bar{G}_{+\alpha,-\alpha}(E)=-U_{\alpha}(-E)^*P_-\;,\nonumber\\
\end{eqnarray}
and
\begin{eqnarray}
&&\!\!\!\!\!\!\!\!\!\!\!\!\!\!\!
\bar{G}_{+\alpha}(E) P_+(E_1)\bar{G}_{-\alpha'}(E')=\phantom{\frac{1}{2}}\nonumber\\
&&t_\alpha(E)\left( \begin{array}{cccc}
t_\alpha^2(-E')^*&t_\alpha^3(-E')^* &0&0 \\
 -t_\alpha^2(-E')^*&-t_\alpha^3(-E')^*&0&0 \\
 0&0&0&0\\
0&0  &0&0\end{array} \right)\;,
\end{eqnarray}
with 
\begin{equation}
\bar{G}_{+\alpha}(E)
P_+(E_1)\bar{G}_{-\alpha'}(E')=-\bar{G}_{-\alpha}(-E)^*P_-\bar{G}_{\alpha}(-E')^*\;.
\end{equation}

Using Eqs.~(\ref{eq:Pi_approx}) and (\ref{eq:K_function_approx}), the explicit
flow equations for the rates $\Gamma_{\pm}(E)$ and the level position
$\epsilon(E)$ are determined from the above Eq.~(\ref{eq:rgl}) for $L_D$ and
its parametrization (\ref{eq:L_form}) to
\begin{eqnarray}
\nonumber
\frac{d}{d\Lambda}\Gamma_{\pm}(E)&=&
\pm i\sum_{\alpha}\frac{Z(E+\mu_{\alpha})}{\Lambda-i\chi(E+\mu_{\alpha})}
t_\alpha(E)t_{2/3}^{\alpha}(-E-\mu_{\alpha})^* \\
\label{eq:gamma_pm_rg} &&
+\,(E\to-E)^*\;,\phantom{\frac{!}{2}}\\
&&\nonumber\\
\nonumber\frac{d}{d\Lambda}\epsilon(E) &=& 
\sum_{\alpha}\frac{Z_1(E-\mu_{\alpha})}{\Lambda-i\chi_1(E-\mu_{\alpha})}
t_\alpha(E-\mu_{\alpha})t_\alpha^\prime(E-\mu_{\alpha}) \nonumber\\
&&\qquad\!\!-2iZ(E)\,\gamma(E)\,{\rm ln} \frac{2\Lambda -i\chi(E)}{\Lambda -i\chi(E)}\;,
\end{eqnarray}
with $\gamma(E)=\sum_{\alpha}U_{\alpha}(E)^2$ and 
$t_\alpha^\prime(E)=t_\alpha^2(-E-\mu_\alpha)^* -t_\alpha^3(-E-\mu_\alpha)^*$.
Similarly Eq.~(\ref{eq:rgvert1}) is evaluated using matrix (\ref{eq:vert1}) to
\begin{eqnarray}
\nonumber
\frac{d}{d\Lambda}t_\alpha(E) &=&
-\frac{Z(E+\mu_{\alpha})}{\Lambda-i\chi(E+\mu_{\alpha})}U_{\alpha}(E+\mu_{\alpha})t_\alpha(E)\;,\\
\nonumber
\frac{d}{d\Lambda}t_\alpha^{2/3}(E) &=& 
-\frac{Z(-E)^*}{\Lambda+i\chi(-E)^*}U_{\alpha}(-E)^*t_\alpha^{2/3}(E)\;,
\end{eqnarray}
which yields
\begin{eqnarray}
\nonumber
t_\alpha(E)\! &=& \frac{1}{2}[t_\alpha^2(-E-\mu_{\alpha})^*\!+t_\alpha^3(-E-\mu_{\alpha})^*]\,,\\ 
\nonumber
t_\alpha^\prime(E) \!&=& t_\alpha^2(-E-\mu_{\alpha})^*\!-t_\alpha^3(-E-\mu_{\alpha})^* = 2\pi i U^{(0)}_{\alpha}t_\alpha(E)\;,
\end{eqnarray}
since the corresponding flow equations have the same form and the initial
conditions are equal.  As a consequence, introducing the rates
$\Gamma(E)=\Gamma_+(E)+\Gamma_-(E)$ and
$\Gamma'(E)=(\Gamma_+(E)-\Gamma_-(E))/2$, the flow equations for the two rates
can be expressed in terms of the single hopping variable $t_\alpha(E)$.  The
flow equation for $U_{\alpha}$ is obtained by using Eq.~(\ref{eq:vert2}) and
integrating Eq.~(\ref{eq:rgvert2}) over $\omega_3$
\begin{equation}
\frac{d}{d\Lambda}U_{\alpha}(E)=
-\frac{2\Lambda\, Z(E)^2}{(\Lambda-i\chi(E))(2\Lambda-i\chi(E))}U_{\alpha}(E)\,\gamma(E)\;.
\end{equation}
Eqs.~(\ref{eq:ivert}) and (\ref{eq:ikernel}), together with the RG
equation for the Liouvillian (\ref{eq:rgl}), yield
\begin{equation}
\nonumber
\frac{d}{d\Lambda}t_{\alpha}^{\gamma}(E) =
-\frac{Z(E+\mu_{\alpha})}{\Lambda-i\chi(E+\mu_{\alpha})}U_{\alpha}(E+\mu_{\alpha})t_{\alpha}^{\gamma}(E)
\end{equation}
for the current hopping amplitude, and
\begin{eqnarray}
\nonumber 
\frac{d}{d\Lambda}\Gamma_{\gamma}^{1/2}(E) &=& 
i\sum_{\alpha}\frac{Z(E+\mu_{\alpha})}{\Lambda-i\chi(E+\mu_{\alpha})}
t_{\alpha}^{\gamma}(E)t_\alpha^{2/3}(-E-\mu_{\alpha})^* \\
\nonumber && +\,(E\to-E)^*
\end{eqnarray}
for the current rates. 
Comparing with the equations for $t_\alpha(E)$ and $\Gamma_{\pm}(E)$ and
considering the respective initial conditions of Sec.~\ref{sec:flow} it follows
\begin{equation}
\label{eq:current_hopping_rate}
t_{\alpha}^{\gamma}(E)=\delta_{\alpha\gamma}t_\alpha(E)\;,\quad
\Gamma_{\pm}(E)=\pm\sum_\alpha\Gamma_{\alpha}^{1/2}(E)\;.
\end{equation}

Summarizing, the flow equations
for the effective model parameters read
\begin{eqnarray}
\frac{d}{d\Lambda}\Gamma_{\alpha}(E)\!&=&\! 
-2\pi\frac{Z(E+\mu_\alpha)}{\Lambda-i\chi(E+\mu_{\alpha})}U^{(0)}_{\alpha}t_\alpha(E)^2\nonumber\\
&&\,-\,(E\to-E)^*\;,\nonumber\\
\frac{d}{d\Lambda}\Gamma'_{\alpha}(E)&=&
i\frac{Z(E+\mu_{\alpha})}{\Lambda-i\chi(E+\mu_{\alpha})}
t_{\alpha}(E)^2+\,(E\to-E)^*\;,\phantom{\sum_{\alpha'}}\nonumber\\
\frac{d}{d\Lambda}t_{\alpha}(E)\!&=&\!
-\frac{Z(E+\mu_{\alpha})}{\Lambda-i\chi(E+\mu_{\alpha})}U_{\alpha}(E+\mu_{\alpha})t_{\alpha}(E)\;,\nonumber\\
\frac{d}{d\Lambda}\epsilon(E)\!&=&
\!2\pi i\sum_{\alpha}\frac{Z_1(E-\mu_{\alpha})}{\Lambda-i\chi_1(E-\mu_{\alpha})}U^{(0)}_{\alpha}
t_{\alpha}(E\!-\!\mu_{\alpha})^2\nonumber\\
&&\!-2iZ(E)\,\gamma(E)\,{\rm ln} \frac{2\Lambda -i\chi(E)}{\Lambda -i\chi(E)}\;,\nonumber\\
\frac{d}{d\Lambda}U_{\alpha}(E)\!&=&\!
-\frac{2\Lambda\,Z(E)^2}{(\Lambda-i\chi(E))(2\Lambda-i\chi(E))}U_{\alpha}(E)\,\gamma(E)\;,\nonumber \\
\label{eq:rg_intermediate}&&
\end{eqnarray}
where we introduced the rates
$\Gamma_\alpha(E)=\Gamma_\alpha^1(E)-\Gamma_\alpha^2(E)$ and
$\Gamma_\alpha^\prime(E)=(1/2)(\Gamma_\alpha^1(E)+\Gamma_\alpha^2(E))$,
according to the definitions \eqref{eq:definitions}.
 
We now determine and discuss the equations for the $Z$-factors $Z(E)$ and
$Z_1(E)$, and subsequently for $\chi(E)$ and $\chi_1(E)$. For $Z(E)$ and
$Z_1(E)$ we find
\begin{eqnarray}
\frac{d}{d\Lambda}Z(E)&=& Z(E)^2\frac{d}{dE}\frac{d}{d\Lambda}\epsilon(E)\;,\nonumber\\
\frac{d}{d\Lambda}Z_1(E)&=& -i Z_1(E)^2\frac{d}{dE}\frac{d}{d\Lambda}\Gamma(E)\;.
\end{eqnarray}
We insert the flow equations (\ref{eq:rg_intermediate}) for $\epsilon(E)$ and
$\Gamma(E)$ and neglect the derivative with respect to $E$ of $Z_i(E)$,
$t_{\alpha}(E)$, and $U_{\alpha}(E)$ on the right-hand side, as their
$E$-dependence is logarithmically weak. This implies
$\frac{d}{dE}\chi_i(E)\simeq Z_i(E)(1-\frac{d}{dE}\lambda_i(E))=1$ and hence
\begin{eqnarray}
\nonumber
\frac{d}{d\Lambda}Z(E) &=&\\
\nonumber
&&\hspace{-1.5cm}
\,-Z(E)^2\left[2\pi\sum_{\alpha}\frac{Z_1(E-\mu_{\alpha})}{(\Lambda-i\chi_1(E-\mu_{\alpha}))^2}
U^{(0)}_{\alpha}t_{\alpha}(E-\mu_{\alpha})^2\;,\right.\\
\nonumber
&&\left.\;-\frac{2\Lambda\,Z(E)}{(\Lambda-i\chi(E))(2\Lambda-i\chi(E))}\gamma(E)\right]\\
&&\nonumber\\
\nonumber
\frac{d}{d\Lambda}Z_1(E) &=&
2\pi Z_1(E)^2 \sum_{\alpha}\frac{Z(E+\mu_{\alpha})}{(\Lambda-i\chi(E+\mu_{\alpha}))^2}
U^{(0)}_{\alpha}t_{\alpha}(E)^2 \\
&&\hspace{2cm}\;
+ \,(E\to-E)^*\;.
\end{eqnarray}
The first terms include an additional factor $t_{\alpha}^2/\Lambda$
and can be neglected to leading order, yielding
\begin{equation}
\label{eq:z}
\frac{d}{d\Lambda}Z(E)\simeq \frac{2 \Lambda\, Z(E)^3\gamma(E)}{(\Lambda-i\chi(E))(2\Lambda-i\chi(E))}
\end{equation}
and a constant for $Z_1(E)\simeq1$. The comparison with the equation for
$U_{\alpha}(E)$ implies
\begin{eqnarray}
\label{eq:U_renorm}
Z(E)U_{\alpha}(E)\simeq U_{\alpha}^{(0)}\;,
\end{eqnarray}
i.e. the product $Z(E)U_{\alpha}(E)$ being unrenormalized.  This simplifies
the flow equation for $t_{\alpha}(E)$ in (\ref{eq:rg_intermediate}) to
\begin{equation}
\label{eq:t_alpha_simplified}
\frac{d}{d\Lambda}t_{\alpha}(E)\!=\!
-\frac{U_\alpha^{(0)}}{\Lambda-i\chi(E+\mu_{\alpha})}t_{\alpha}(E)\;.
\end{equation}

We can now derive the equation for
$\chi(E)=Z(E)(E-\epsilon(E))$.  The above expressions yield
\begin{eqnarray}
&&\!\!\,\frac{d}{d\Lambda}Z(E)\epsilon(E)= 
\nonumber\\&&\;\;\;2\pi i\sum_{\alpha}\frac{Z(E)}{\Lambda-i\chi_1(E-\mu_{\alpha})}
U_{\alpha}^{(0)}t_{\alpha}(E-\mu_{\alpha})^2\nonumber\\
&&\;\;\;-2\gamma_0\left[i\,{\rm ln}\frac{2\Lambda-i\chi(E)}{\Lambda-i\chi(E)}
-\frac{\Lambda\,Z(E)\epsilon(E)}{(2\Lambda-i\chi(E))(\Lambda-i\chi(E))}\right]\;,\nonumber\\&&
\end{eqnarray}
with $\chi_1(E)=E+i\Gamma(E)$ and $\gamma_0=\sum_\alpha (U_\alpha^{(0)})^2$.  
From Eq. (\ref{eq:z}) for $Z(E)$ the equation for $\chi(E)$ reads
\begin{eqnarray}
&&\!\!\frac{d}{d\Lambda}\chi(E)=
\nonumber\\&&\;\;\;-2\pi i\sum_{\alpha}\frac{Z(E)}{\Lambda-i\chi_1(E-\mu_{\alpha})}
U_{\alpha}^{(0)}t_{\alpha}(E-\mu_{\alpha})^2\nonumber\\
&&\;\;\;+2\gamma_0\left[i\,{\rm ln}\frac{2\Lambda-i\chi(E)}{\Lambda-i\chi(E)}
+\frac{\Lambda\chi(E)}{(2\Lambda-i\chi(E))(\Lambda-i\chi(E))}\right]\;.\nonumber\\&&
\end{eqnarray}
In leading order, this equation can approximately be integrated by
\begin{eqnarray}
\chi(E)&=&2i\gamma_0\Lambda\,{\rm ln}\,\frac{2\Lambda-i\chi(E)}{\Lambda-i\chi(E)}+\chi'(E)\;,
\label{eq:chi}
\end{eqnarray}
with 
\begin{equation}
\label{eq:chi_prime_rg}
\frac{d}{d\Lambda}\chi'(E)= -2\pi i\sum_{\alpha}\frac{Z(E)}{\Lambda-i\chi_1(E-\mu_{\alpha})}
U_{\alpha}^{(0)}t_{\alpha}(E-\mu_{\alpha})^2\;.
\end{equation}
The terms neglected stem from the $\Lambda$-dependence of $\chi(E)$ leading
to higher order terms of order $\gamma_0{d\over d\Lambda}\chi(E)\sim O(U^3)$.
The first term of \eqref{eq:chi_prime_rg} is important since it cancels the
large term proportional to $D$ in the initial condition for $\epsilon(E)$, 
see \eqref{eq:initial_values}. To achieve this, the following relation is 
needed between the physical reservoir band width and the initial cutoff of 
the RG flow
\begin{equation}
\label{eq:D_initial_cutoff}
\Lambda_0\,=\,{\pi^2\over 16 \ln{2}}D\;.
\end{equation}
Using \eqref{eq:initial_values} and neglecting unimportant terms $\sim
O(\gamma_0)$, the initial condition for $\chi^\prime(E)$ then reads
\begin{equation}
\label{eq:chi_prime_initial}
\chi^\prime(E)|_{\Lambda_0}=E-\epsilon_0+i{\Gamma^{(0)}\over 2}\;.
\end{equation}

The RG equation \eqref{eq:chi_prime_rg} together with the initial condition
\eqref{eq:chi_prime_initial} lead to the RG equation \eqref{eq:rg_chi} with
the definition $\tilde{\Gamma}_\alpha(E)=2\pi Z(E+\mu_\alpha)t_\alpha(E)^2$.
Since the first term of \eqref{eq:chi} is of order $\sim \gamma_0$ it can be
neglected in all denominators of \eqref{eq:rg_intermediate} and
$(\Lambda-i\chi(E))^{-1}\simeq(\Lambda-i\chi'(E))^{-1}$.  Thus, the first two
RG equations of \eqref{eq:rg_intermediate} are identical to the RG equations
\eqref{eq:rg_gamma} and \eqref{eq:rg_gamma_prime}. Finally, one obtains the RG equation
\eqref{eq:rg_gamma_tilde} for $\tilde{\Gamma}_\alpha(E)$ if one
combines the RG equations \eqref{eq:t_alpha_simplified} for $t_\alpha(E)$ with
the RG equation \eqref{eq:z} for $Z(E)$ and uses \eqref{eq:U_renorm}.

\section{Contour integrations for the time evolution}
\label{app:time} 

\begin{figure}[t!]
  \centering
  \includegraphics[width=60mm,clip=true]{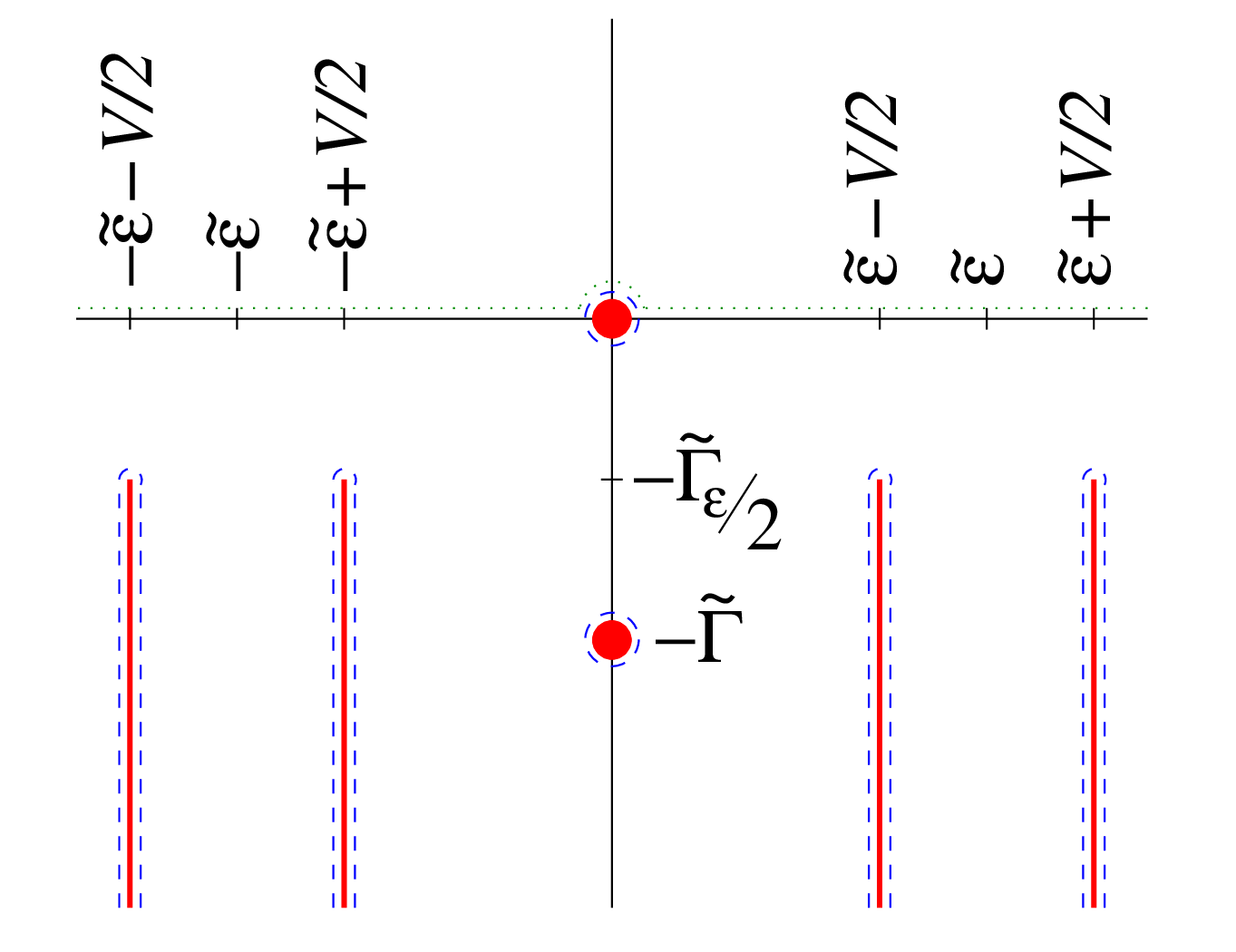}
  \caption{(Color online) Analytic structure of the integrand of
    $J_\pm(t)$. All singularities appear in the lower half plane. Red
    dots stand for poles while red solid lines represent branch cuts.  The
    pole at $z=0$ corresponds to the stationary state. The blue dashed and
    green dotted lines are the original and deformed integration contours
    respectively.}
  \label{fig:fig13}
\end{figure}

The evaluation of the auxiliary functions $J_\pm(t)$ defined in
\eqref{eq:timeint1} is performed using standard techniques for contour
integrations. The integrand has poles at $z = 0$ and $z=-i\tilde{\Gamma}$
as well as branch cuts starting at $z=\tilde{\epsilon}\pm
V/2-i\tilde{\Gamma}_\epsilon/2$ and $z=-\tilde{\epsilon}\pm
V/2-i\tilde{\Gamma}_\epsilon/2$, see Fig.~\ref{fig:fig13}. 
 
Let us first consider the non-interacting case. The solution of RG equations
yields a result for the rates $\Gamma_{\alpha}$ and $\Gamma'_{\alpha}$, which
appears to be exact in the scaling limit. In particular, for symmetric
coupling we obtain $\Gamma (z) = \Gamma_L^{(0)} +\Gamma_R^{(0)} =
\Gamma^{(0)}\equiv T_K$, $\Gamma_{\pm} (z) = \frac12  T_K\pm
\sum_{\alpha} \Gamma'_{\alpha} (z)$, $\Gamma_{\alpha}^{1/2} = \Gamma'_{\alpha}
\pm \frac14  T_K$, and
\begin{eqnarray}
\Gamma'_{\alpha} (z) &=&  \frac{i  T_K}{4 \pi} 
\left[ \ln \left( \frac{ T_K}{2} - i (z +\mu_{\alpha} - \epsilon_0 ) \right) 
\right. \nonumber \\
& & \left. \quad - \ln \left( \frac{ T_K}{2} 
- i (z -\mu_{\alpha} + \epsilon_0) \right) \right].
\label{eq:gamma_prime_NI}
\end{eqnarray}

Using these values we obtain \eqref{eq:n_NI} and \eqref{eq:I_NI} from
\eqref{eq:dot_occup_int} and \eqref{eq:timeint2}, respectively, where 
the functions $ F_{0/1, \alpha} (t)$ are originally defined by
\begin{equation}
F_{0/1,\alpha} (t)= \frac{i}{2 \pi  T_K} 
\int_{-\infty+i 0^+}^{\infty + i 0^+} d z e^{-i z t} 
\frac{\Gamma'_{\alpha} (z)}{z + i \frac{ T_K}{2} \mp i \frac{ T_K}{2}}\;.
\label{eq:F01}
\end{equation}
We note their important property
\begin{equation}
\frac{d}{d t} \left[ F_{0, \alpha} (t) - F_{1,\alpha} (t) \right] = 
T_K F_{1,\alpha} (t)\;,
\end{equation}
which guarantees the current conservation $\frac{d }{d t} n (t) =
\sum_{\gamma} I_{\gamma} (t)$.

Let us now show that the result of integration in \eqref{eq:F01} leads to
Eq.~\eqref{eq:F01main}. To this end we deform the contour of integration from
the real axis to the paths embracing the poles and the branch cuts shown in
the Fig.~\ref{fig:fig13}. The positions of the nonzero pole as well as of the
branch points of $\Gamma'_{\alpha} (z)$ (see Eq.~\eqref{eq:gamma_prime_NI}
above) in the integrand of \eqref{eq:F01} are given by the bare values of
$\tilde{\Gamma} = \tilde{\Gamma}_{\epsilon} = T_K$ and $\tilde{\epsilon} =
\epsilon_0$. We obtain the following contributions to $F_{0/1,\alpha} (t) =
F^p_{0/1, \alpha} (t) + F^{br.c.}_{0/1, \alpha} (t)$. The pole contribution
equals
\begin{equation}
F^p_{0, \alpha} (t) = \frac{\Gamma'_{\alpha} (0)}{T_K}, 
\quad F^p_{1, \alpha} (t) = 
\frac{\Gamma'_{\alpha} (- i T_K )}{T_K} e^{-T_K t} ,
\label{eq:F01p}
\end{equation}
where
\begin{eqnarray}
\Gamma'_{\alpha} (0)& =& - \Gamma'_{\alpha} (-i T_K)+ \frac{ T_K}{2} \mathrm{sign}(\mu_{\alpha} - \epsilon_0)\\
&=& 
\frac{T_K}{2 \pi} \arctan \frac{\mu_{\alpha} - \epsilon_0}{T_K/2}+ \frac{ T_K}{2} \mathrm{sign}(\mu_{\alpha} - \epsilon_0)\,. \nonumber
\label{eq:gam_prime_pole}
\end{eqnarray}
The branch-cut contribution equals
\begin{eqnarray}
F_{0/1, \alpha}^{br.c.} (t) &=& 
\frac{i e^{i (\epsilon_0 - \mu_{\alpha}) t}}{2 \pi T_K} 
\int_{- i \infty}^{- i \frac{T_K}{2}}  
\frac{d (i y) e^{y t}}{i y + i \frac{T_K}{2} \mp i \frac{T_K}{2}  
- (\epsilon_0 - \mu_{\alpha})} \nonumber \\
& & \times \frac{i T_K}{4 \pi} (- 2 \pi i)  - 
(\epsilon_0 - \mu_{\alpha} \to - \epsilon_0 + \mu_{\alpha}) \nonumber \\
&=& - \mathrm{Im} \, \frac{ e^{i (\epsilon_0 - \mu_{\alpha}) t - 
\frac{T_K t}{2}}}{2 \pi} \int_{-  \infty}^{0}  
\frac{d y \, e^{y t}}{y  \mp  \frac{T_K}{2} + 
i (\epsilon_0 - \mu_{\alpha})}  \nonumber \\
&=&  \mathrm{Im} \, \frac{e^{i (\epsilon_0 - \mu_{\alpha}) t 
- \frac{T_K t}{2}}}{2 \pi }  \int_0^{\infty} \frac{d x \, e^{-x}}
{x \pm \frac{T_K t}{2} - i (\epsilon_0 - \mu_{\alpha}) t}  \nonumber \\
&=& - \frac{e^{-T_K t/2 \pm T_K t/2}}{2 \pi} \, \mathrm{Im} \, 
\mathrm{Ei} \left(\mp \frac{T_K t}{2} + i (\epsilon_0 - \mu_{\alpha}) 
t \right),
\nonumber \\
\label{eq:F01brc}
\end{eqnarray}
where we exploit the analytic continuation of the exponential integral
function (see 8.212.5 of Ref.~\onlinecite{GR} and
Ref.~\onlinecite{AbramowitzStegun})
\begin{equation}
\mathrm{Ei} (\pm z) = - e^{\pm z} \int_0^{\infty} 
\frac{e^{- x}}{x \mp z} d x, \quad (\mathrm{Re} \, z >0 ) .
\end{equation}
Combining \eqref{eq:F01p}, \eqref{eq:gam_prime_pole} and \eqref{eq:F01brc} we
obtain the formula \eqref{eq:F01main}. We also note that
\begin{eqnarray}
F_{0/1, \alpha}^{br.c.} (t=0^+) &=&  - \frac{1}{2 \pi} \mathrm{Im} \,  
\int_{-  \infty}^{0} \frac{d y}{y  \mp  \frac{T_K}{2} + 
i (\epsilon_0 - \mu_{\alpha})} 
\nonumber \\
&=& \pm \frac{1}{2 \pi}  \arctan \frac{\epsilon_0 - \mu_{\alpha}}{T_K/2} ,
\end{eqnarray}
which implies the property
\begin{equation}
F_{0/1,\alpha} (t=0^+)  =0.
\label{eq:F01_0}
\end{equation}

In the interacting case, the analytic structure remain very similar to that of
the non-inteacting case. The main difference is contained in the type of
branching behavior which changes from the logarithmic to the power-law one.
Additionally, positions of the branch point as well as a position of the
nonzero pole are shifted to interaction-dependent values.

\begin{figure}[t]
  \centering
  \includegraphics[width=60mm,clip=true]{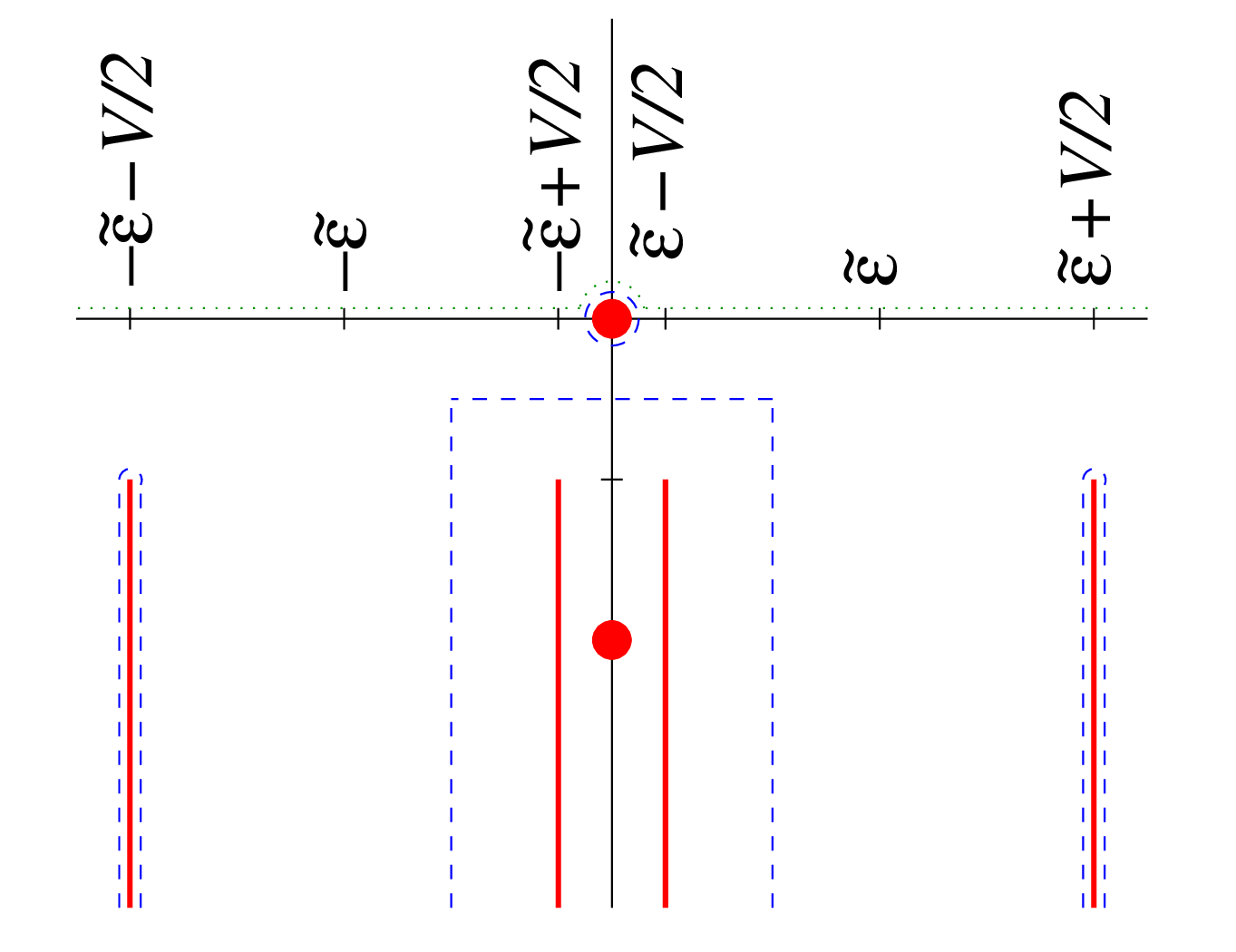}
  \caption{(Color online) Integration contour for $J_\pm(t)$ 
    on resonance $|\tilde{\epsilon} -V/2|\ll T_K$.}
  \label{fig:fig14}
\end{figure}

For $|\tilde{\epsilon}-V/2|\gg T_K$ we can treat all poles and branch
cuts separately. Thus the evaluation of $J_\pm(t)$ boils down to
\begin{eqnarray}
  J_\pm(t)\!\!&=&\!\!-\frac{1}{2}\mp\frac{\Gamma'(0)}{\Gamma(0)}\\
  &&\!\!+\!\!\left(\frac{1}{2}\pm
    \frac{\Gamma'(-i\tilde{\Gamma})}{\tilde{\Gamma}}\right)
  e^{-\tilde{\Gamma}t}\,
  \text{Res}
  \left(\frac{1}{z+i\Gamma(z)},z=-i\tilde{\Gamma}\right)\nonumber\\
  & &\!\!+\frac{i}{2\pi}\!\sum_{\alpha=\pm V/2}
  \sum_{\beta=\pm\tilde{\epsilon}}e^{-i(\alpha+\beta)t}
  \int_{-\infty}^{-\tilde{\Gamma}_\epsilon/2}
  \frac{dy\,e^{yt}}{\alpha+\beta+i y} \nonumber\\*
  &&\!\!\times\!\!\left(\frac{\Gamma_\pm(\alpha+\beta-\eta+i y)}
    {\alpha+\beta+i y+i\Gamma(\alpha+\beta-\eta+i y)}
    -(\eta\rightarrow -\eta)\right)\;.\nonumber
\end{eqnarray}
Using $\Gamma_\pm(z)=\frac{1}{2}\Gamma(z)\pm\Gamma'(z)$ it can be
cast in the form
\begin{eqnarray}
\!\!\! J_\pm(t)&=&-\frac{1}{2}\mp\frac{\Gamma'(0)}{\Gamma(0)}\\
&&+\left(\frac{1}{2}\pm
    \frac{\Gamma'(-i\tilde{\Gamma})}{\tilde{\Gamma}}\right)
  e^{-\tilde{\Gamma}t}\,+J_b(t)\pm J_b'(t)\;,\;\;\;\nonumber
\end{eqnarray}
where the first two terms equal $n^{st}$. Then a straightforward calculation
yields
\begin{widetext}
\begin{eqnarray}
  J_b(t)&=&
  \frac{2\sin(\pi g)}{\pi}
  \frac{e^{-\frac{\tilde{\Gamma}_\epsilon}{2}t}}{(tT_K)^{1+g}}
  \sum_{\alpha=\pm V/2}\sum_{\beta=\pm\tilde{\epsilon}}e^{-i(\alpha+\beta)t}\\*
  & &\times\int_0^\infty\frac{ds}{s^g}\frac{e^{-s}}
  {\left[
      4\frac{(\alpha+\beta-i\tilde{\Gamma}_\epsilon/2)t-i s}{(tT_K)^{1+g}}+
      \frac{i}{(-s-2i\alpha t)^g}+\frac{i}{(-s-2i\beta t)^g}+
      \frac{i}{(-s-2i(\alpha+\beta)t)^g}+\frac{i}{s^g}\cos(\pi g)\right]^2
    +\frac{\sin^2(\pi g)}{s^{2g}}}\;,
  \nonumber
\end{eqnarray}
\end{widetext}
while $J_b'(t)$ has been evaluated numerically.
We note that the factor $\beta/\tilde{\epsilon}$ just gives a sign. For
$V,|\tilde{\epsilon} -V/2|\gg T_K,1/t$ we obtain \eqref{eq:ntresult}. Close to
resonance, $|\tilde{\epsilon} - V/2|\ll T_K$, the two branch cuts starting at
$z=\pm\tilde{\epsilon}\mp V/2-i\tilde{\Gamma}_\epsilon/2$ are very close to
the pole at $z=-i\tilde{\Gamma}$, which leads to a numerical instability in
the calculation of its residue. Therefore, it is advantageous to directly
evaluate $J_\pm(t)$ as defined in \eqref{eq:timeint1} on the contour shown in
Fig.~\ref{fig:fig14}, thereby encircling the pole at $z=-i\tilde{\Gamma}$.

The time evolution of the current given in \eqref{eq:timeint2} can be
cast in the form
\begin{equation}
  I_L(t)=J^1(t)+\left(\frac{1}{2}-n(0)\right)J^2(t)+J^3(t)\;,
\end{equation}
where
\begin{eqnarray}
  J^1(t)&=&\frac{i}{2\pi}\int_{-\infty+i 0^+}^{\infty+i 0^+}
  \frac{dz}{z}\,\Gamma_L'(z)\,e^{-i zt}\;,\\
  J^2(t)&=&\frac{i}{2\pi}\int_{-\infty+i 0^+}^{\infty+i 0^+}
  \frac{dz}{z+i\Gamma(z)}\,\Gamma_L(z)\,e^{-i zt}\;,\\
  J^3(t)&=&\frac{1}{2\pi}\int_{-\infty+i 0^+}^{\infty+i 0^+}
  \frac{dz}{z}\,\frac{\Gamma'(z)\,\Gamma_L(z)}{z+i\Gamma(z)}\,e^{-i zt}\;.
\end{eqnarray}
The evaluation is analogous to the one for $J_\pm(t)$
above. In particular, in the long-time limit
$Vt,|\tilde{\epsilon}-V/2|t\gg 1$ off resonance we find
\begin{eqnarray}
  J^1(t)\!\!&=&\!\!\Gamma_L'(z=0)+\frac{T_K}{2\pi}\,(T_Kt)^g\,
  e^{-\tilde{\Gamma}_\epsilon t/2}\,
  \frac{\cos\bigl((\tilde{\epsilon}-V/2)t\bigr)}
  {(\tilde{\epsilon}-V/2)t}\;,\nonumber\\
  J^2(t)\!\!&=&\!\!\Gamma_L(-i\tilde{\Gamma})\,
  e^{-i\tilde{\Gamma}t}\;,\nonumber\\
  J^3(t)\!\!&=&\!\!-\frac{\Gamma'(0)}{\Gamma(0)}\Gamma_L(0)+
  \frac{\Gamma'(-i\tilde{\Gamma})}{\tilde{\Gamma}}\,
  \Gamma_L(-i\tilde{\Gamma})\,e^{-i\tilde{\Gamma}t}\;,
\end{eqnarray}
where for $J^2$ and $J^3$ terms in $O(U^{(0)})$ were neglected. The first
terms in $J^1$ and $J^3$ together yield the stationary current
\eqref{eq:I_stationary}.


\vfill\eject


\begin{thebibliography}{99}

\bibitem{VigmanFinkelstein78}
P.B. Vigman and A.M. Finkel'shtein, Sov. Phys. JETP \textbf{48}, 102 (1978).

\bibitem{Schlottmann}
P. Schlottmann, J. Magn. magn. Mater. \textbf{7}, 72 (1978); J. Phys. (Paris)
\textbf{39}, C6-1486 (1978).

\bibitem{Anderson}
P.W. Anderson and G. Yuval, Phys. Rev. Lett. \textbf{23}, 89 (1969);
G. Yuval and P.W. Anderson, Phys. Rev. B \textbf{1}, 1522 (1970); 
P.W. Anderson, G. Yuval, and D.R. Hamann, Phys. Rev. B \textbf{1}, 
4464 (1970);
K.D. Schotte, Z. Physik \textbf{230}, 99 (1970).

\bibitem{Toulouse}
G. Toulouse, C. R. Acad. Sci. (Paris) \textbf{268}, 1200 (1969).

\bibitem{FilyovWiegmann80}
V.M. Filyov and P.B. Wiegmann, Phys. Lett.~A \textbf{76}, 283 (1980);
A.M. Tsvelick and P.B. Wiegmann, Adv. Phys. \textbf{32}, 453 (1983).

\bibitem{Schlottmann82} P. Schlottmann, Phys. Rev.~B \textbf{25}, 4815 (1982);
  \emph{ibid.} \textbf{25}, 4828 (1982); \emph{ibid.} \textbf{25}, 4838
  (1982).

\bibitem{GNT}
A.O. Gogolin, A.A. Nersesyan, and A.M. Tsvelik, \emph{Bosonization and
  Strongly Correlated Systems} (Cambridge University Press, Cambridge, 2004).

\bibitem{MehtaAndrei06}
P.~Mehta and N.~Andrei, Phys.~Rev.~Lett.~\textbf{96}, 216802 (2006); 
P.~Mehta, S.-P. Chao, and N.~Andrei, arXiv:cond-mat/0703426.

\bibitem{Doyon}
B.~Doyon, Phys.~Rev.~Lett. {\bf 99}, 076806 (2007).

\bibitem{Golub}
A. Golub, Phys.~Rev.~B~\textbf{76}, 193307 (2007).

\bibitem{NH} 
A. Nishino and N. Hatano, J.~Phys.~Soc.~Jpn. {\bf 76}, 063002 (2007).

\bibitem{NIH} A. Nishino, T. Imamura, and N. Hatano, Phys.~Rev.~Lett. {\bf
    102}, 146803 (2009).

\bibitem{BS}
E.~Boulat and H.~Saleur, Phys.~Rev.~B~\textbf{77}, 033409 (2008). 

\bibitem{BSS} E.~Boulat, H. Saleur, and P. Schmitteckert, Phys.~Rev.~Lett.
  {\bf 101}, 140601 (2008).
  
\bibitem{BBSS} A. Bransch\"adel, E.~Boulat, H. Saleur, and P. Schmitteckert,
  Phys. Rev. Lett. \textbf{105}, 146805 (2010).

\bibitem{BVZ} L.~Borda, K. Vlad\'{a}r, and A. Zawadowski, Phys. Rev. B
  \textbf{75}, 125107 (2007);
  L.~Borda, A. Schiller, and A. Zawadowski, Phys.  Rev. B \textbf{78},
  201301(R) (2008).

\bibitem{karrasch10} 
C. Karrasch, M. Pletyukhov, L. Borda, and V.  Meden, 
Phys. Rev. B \textbf{81}, 125122 (2010).

\bibitem{SA}
C. Karrasch, S. Andergassen, M. Pletyukhov, D. Schuricht,
L. Borda, V. Meden, and H. Schoeller, Europhys. Lett. \textbf{90},
30003 (2010).

\bibitem{jakobs07}
S.G.~Jakobs, V.~Meden, and H.~Schoeller,
Phys.~Rev.~Lett.~{\bf 99}, 150603 (2007);
S.G.~Jakobs, M.~Pletyukhov, and H.~Schoeller,
Phys.~Rev.~B {\bf 81}, 195109 (2010).
R.~Gezzi, Th. Pruschke, and V. Meden, Phys.~Rev.~B {\bf 75}, 045324 (2007).

\bibitem{S}
H. Schoeller, Eur. Phys. J. Special Topics {\bf 168}, 179 (2009).

\bibitem{review}
S. Andergassen, V. Meden, H. Schoeller, J. Splettstoesser, and M.R. Wegewijs,
Nanotechnology {\bf 21}, 272001 (2010).

\bibitem{RTRGKondo}
H. Schoeller and F. Reininghaus, Phys. Rev. B {\bf 80}, 045117 (2009);
\emph{ibid.} Phys.~Rev.~B {\bf 80}, 209901(E) (2009);
D. Schuricht and H. Schoeller, Phys. Rev. B {\bf 80}, 075120 (2009).

\bibitem{PSS} M. Pletyukhov, D. Schuricht, and H. Schoeller,
  Phys. Rev. Lett. \textbf{104}, 106801 (2010).

\bibitem{DoyonAndrei06}
B. Doyon and N. Andrei, Phys. Rev.~B {\bf 73},  245326  (2006).

\bibitem{Kehrein05}
S. Kehrein, Phys. Rev. Lett. {\bf 95}, 056602 (2005); 
S. Kehrein, {\em The Flow Equation Approach to Many-Particle Systems}
  (Springer, Berlin, 2006); 
P. Fritsch and S. Kehrein, Ann. Phys. {\bf 324}, 1105 (2009); 
P. Fritsch and S. Kehrein, Phys. Rev.~B {\bf 81}, 035113 (2010).

\bibitem{ndc}
L. Borda and A. Zawadowski, Phys. Rev. B {\bf 81}, 153303 (2010).

\bibitem{endnote3} This corrects a typographical error in Eq. (8) of
  Ref.~\onlinecite{SA}. 

\bibitem{ECS}
J. Peskill, in \emph{Introduction to Quantum Computation and Information}
(H.-K. Lo, S. Popescu, and T. Spiller, World Scientific, Singapore, 1998);
D.P. DiVincenzo and D. Loss, Phys.~Rev.~B \textbf{71}, 035318 (2005);
J. Fischer and D. Loss, Science \textbf{324}, 1277 (2009).

\bibitem{displacementI} M. B\"uttiker, A. Pr\^{e}tre, and H. Thomas,
  Phys. Rev. Lett. \textbf{70}, 4114 (1993); J. Fransson, O. Eriksson,
  and I. Sandalov, Phys. Rev.~B \textbf{66}, 195319 (2002).

\bibitem{Schmidtea} T.L.~Schmidt, P. Werner, L. M\"uhlbacher, and
  A. Komnik, Phys.~Rev.~B {\bf 78}, 235110 (2008).

\bibitem{Leggett}
A.J. Leggett, S.~Chakravarty, A.T. Dorsey, M.P.A. Fisher, A.~Garg, and
  W.~Zwerger, Rev. Mod. Phys. \textbf{59},  1 (1987).

\bibitem{LesageSaleur}
F.~Lesage and H.~Saleur, Phys.~Rev.~Lett.~{\bf 80}, 4370 (1998).

\bibitem{AndersSchiller}
F.B.~Anders and A.~Schiller, Phys.~Rev.~Lett.~{\bf 95}, 196801 (2005);
Phys.~Rev.~B~{\bf 74}, 245113 (2006).

\bibitem{Komnik}
A.~Komnik, Phys.~Rev.~B~{\bf 79}, 245102 (2009).

\bibitem{HeylKehrein}
D.~Lobaskin and S.~Kehrein, Phys.~Rev.~B~{\bf 71}, 193303 (2005);
D.~Lobaskin and S.~Kehrein, J. Stat. Phys. \textbf{123}, 301 (2006);
M. Heyl and S.~Kehrein, J. Phys.: Condens. Matter \textbf{22}, 345604 (2010).

\bibitem{GR}
I.S. Gradshteyn and I.M. Ryzhik, {\em Table of Integrals, Series, and
  Products} (Academic Press, London, 1994).

\bibitem{AbramowitzStegun}
M. Abramowitz and I.A. Stegun, \emph{Handbook of Mathematical Functions}
(Dover, New York, 1965).

\end{thebibliography}
\end{document}